\documentclass{arxiv-template}
\usepackage{ragged2e}
\usepackage{indentfirst}
\usepackage{amsmath,amssymb}
\usepackage{booktabs,array}
\usepackage{graphicx}
\usepackage{siunitx}
\usepackage{ifthen}
\usepackage{float}
\usepackage{pdfpages}
\begin{document}
\sloppy
\pagestyle{fancy}
\rhead{}

\title{Kirigami Meta-Sheet for Enhanced Impact Absorption}

\maketitle

% Author: Please give full first and last names for authors and include * after the name of all corresponding authors
\author{Dahyun Joo, Yasuhiro Miyazawa, Seokhyeon Hong, Dongha Lee, Jinkyu Yang,}
\author{Do-Nyun Kim*}

% Dedication
%\dedication{Optional dedication here. If no dedication is required, please leave blank}

% Affiliations: Please provide academic titles (Prof. or Dr.) for all authors where applicable, and include an institutional email address for all corresponding authors
\begin{affiliations} Dahyun Joo, Yasuhiro Miyazawa, Seokhyeon Hong, Dongha Lee, Prof. Jinkyu Yang, Prof. Do-Nyun Kim\\ Department of Mechanical Engineering, Seoul National University, 1 Gwanak-ro, Gwanak-gu, Seoul 08826, Republic of Korea\\ Prof. Do-Nyun Kim\\ Institute of Advanced Machines and Design, Seoul National University, Seoul 08826, Republic of Korea\\ Institute of Engineering Research, Seoul National University, Seoul 08826, Republic of Korea\\ Email Address: dnkim@snu.ac.kr
%A. N. O. Author\\
%Address
\end{affiliations}

% Keywords: Please provide a minimum of three and a maximum of seven keywords, separated by commas

\justifying

% Abstract should be written in the present tense and impersonal style (i.e., avoid we), and be at most 200 words long

\begin{abstract}
Impact absorbers based on mechanical metamaterials often use bulky, vertically stacked architectures, limiting large area deployment and scalable manufacturing. Here, we propose a kirigami meta-sheet as a planar absorber that uses transitions between positive and negative stiffness regimes rather than sacrificial crushing. Guided by an analysis of a simple mass-spring-damper model, we program the stiffness of kirigami meta-sheets through the hinge ratio connecting the unit cells. Quasistatic indentation experiments confirm that the meta-sheet with a low hinge ratio most clearly exhibits the negative stiffness transition. The drop-tower tests show that it reduces rebound, decreases the first impact force, and increases dissipation. Unlike polyethylene mesh and styrofoam, this kirigami meta-sheet is shown to be effective in protecting a falling egg. Its planar geometry enables area scaling by tiling and is compatible with sheet level manufacturing routes such as cutting, molding, and lamination, establishing kirigami meta-sheets as practical impact absorbers.
\end{abstract}

\setlength{\parindent}{2em}

\section{Introduction}
Impact protective materials and structures are widely used to reduce sudden mechanical shocks in packaging, transportation, wearable protection, automotive components, marine structures, and aerospace systems [1,2]. A successful absorber must simultaneously limit peak force, provide a sufficiently long working distance, and dissipate or store incoming kinetic energy without transmitting damaging loads to the protected object. Conventional absorbers achieve this function through plastic deformation, crushing, fracture, fragmentation, dry friction, or rate dependent viscous processes [3--5]. Styrofoam and related foams are representative examples [6,7]: they can effectively attenuate impact, but their protective function is commonly accompanied by irreversible crushing or fracture, making repeated use difficult. Reusable impact absorption therefore requires a structure that can access a large deformation pathway, dissipate energy during loading--unloading, and recover after impact without relying on sacrificial failure [8].

Mechanical metamaterials offer a geometry-based route to this target [9--12]. Their unusual responses originate from architecture rather than chemical composition, allowing lightweight lattices, recoverable buckling, negative stiffness, auxetic deformation [13--15], and programmable force--displacement curves [16--18]. Buckling based metamaterial shock absorbers are particularly relevant because they can exhibit a negative differential stiffness region followed by stiffening, thus producing a long working distance and repeatable energy absorption without permanent material crushing [19--28]. For example, a vertically organized buckling microlattice absorber demonstrates this reusable metamaterial strategy [19].

Despite these advances, most reusable mechanical metamaterial absorbers remain organized as vertical stacks, compression columns, or bulky three dimensional assemblies. These geometries are effective when a protected object can be supported on a thick layer with sufficient vertical clearance, but they are less suitable for broad, thin, fabric like, or surface mounted protection. They also often require support from a ground or backing surface to impose the intended compression boundary condition, so their use is limited to situations where that boundary condition can be provided. A planar absorber can broaden the application space by spreading a local impact into lateral deformation of surrounding units while being held mainly along the edge. If a periodic sheet is sufficiently connected, the same stiffness transition mechanism can be activated over a broad in-plane region, so that impact protection is less sensitive to the exact impact location except near free edges.

Kirigami metamaterials provide a natural basis for such a planar absorber [29--34]. Patterned cuts and compliant hinges allow flat sheets to rotate, buckle out of plane, stretch to large deformations [35], display auxetic response [36], and program stiffness through geometry [37--42]. Previous work on bistable kirigami surfaces showed that hinge geometry can control nonlinear stiffness and the transition between mechanically distinct deformation regimes [43], and related slit patterned thin wall structures further show that cut geometry can redirect impact deformation [44]. Building on this hinge controlled stiffness concept, the present study focuses specifically on the transition between positive stiffness and negative stiffness regions as the primary impact dissipation mechanism. The key unresolved question is whether this stiffness regime transition can be distributed across a planar sheet like architecture and translated into measurable impact energy attenuation.

Here, a horizontally constrained kirigami surface, meta-sheet, is introduced as a planar impact absorber (Figure~1a). It converts local out-of-plane impact into hinge buckling, in-plane unit rotation, lateral force redistribution, and transition induced vibration caused by crossing between positive and negative stiffness regimes. This planar format can be enlarged by tiling the unit pattern and is compatible with sheet level manufacturing routes such as laser cutting, casting, molding, and patterned lamination once the target material system is selected. The study first uses a simple mass-spring-damper model to show why a negative stiffness response improves impact mitigation. Guided by this result, quasistatic indentation experiments identify the nonlinear restoring law, stiffness transition signature, and hysteresis of low hinge ratio (LHR), medium hinge ratio (MHR), and high hinge ratio (HHR) meta-sheet arrays. Finite element (FE) simulations are then used to inspect deformation modes and hinge localized stress concentration that are difficult to isolate experimentally. Drop-tower impact tests connect the measured nonlinear stiffness to rebound reduction, first impact force, and dynamic dissipation. The results demonstrate that planar kirigami can operate as a scalable and mechanically efficient meta-sheet for impact absorption.

\section{Results and Discussion}

\subsection{Kirigami Meta-sheet Design}
We consider a drop impact situation in which an impactor released from height \(h\) accelerates under gravitational acceleration \(\vec{g}\) before contacting the sheet (Figure~1a,b). The proposed meta-sheet design translates the buckling based negative stiffness concept into a planar kirigami meta-sheet with lateral load redistribution through edge constraints. Local out-of-plane impact drives hinge buckling, in-plane unit rotation, and progressive lateral spreading from the impact point to neighboring kirigami units. The intended role of the geometry is therefore not simply to provide a long vertical compression stroke. Instead, a local impact induces stiffness regime transitions progressively across multiple kirigami units over an interior surface area, which is the key distinction from vertically stacked metamaterial absorbers or thick backing dependent structures.

The elementary response is described by three stiffness regimes (Figure~1c). In region~I, the structure is in the regime of positive bending stiffness, where the force increases with displacement as the hinges and surrounding units store elastic energy. Region~II is the regime of negative stiffness, where hinge buckling reduces the restoring force despite continued displacement. The force increases again in region~III due to geometric constraint and large deformation stiffening. This sequence suggests a useful impact mitigation pathway: the structure can first carry the incoming load, then reduce the transmitted force as it enters the negative stiffness region, and finally recover force capacity in the hardening region before excessive collapse occurs.

The stiffness of the hinge is crucial to embed this unique stiffness transition into the meta-sheet (Figure~1d). We define the hinge ratio (HR) as
\begin{equation}
\mathrm{HR}=\frac{t_h}{L_{\mathrm{uc}}},
\label{eq:HR}
\end{equation}
where \(t_h\) is the thickness of the hinge and \(L_{\mathrm{uc}}\) is the effective unit cell length [43]. For the present kirigami pattern, \(L_{\mathrm{uc}}=L/8\) because eight unit cells span the longest dimension \(L\). A smaller HR makes the hinge more compliant and allows access to the negative stiffness region, whereas a larger HR suppresses the force reduction and shifts the response toward positive stiffness dominant behavior.

To illustrate, we can consider three representative cases: low, medium, and high HR values (LHR, MHR, and HHR, respectively). The LHR meta-sheet may exhibit a positive-negative-positive stiffness transition, the MHR may show zero stiffness in region~II, and the HHR may possess a positive stiffness only without a stiffness transition. The corresponding LHR, MHR, and HHR geometries are shown as representative designs (Figure~1e). The planar kirigami units are periodically tiled and mechanically connected, so the same sequence of local indentation, hinge buckling, lateral spreading, stiffness regime transition, and vibration can be activated over the interior of the sheet. The response can therefore be scaled by enlarging or tiling the planar pattern rather than by stacking many vertical units.

\subsection{Effect of Stiffness Profiles on Impact Mitigation}

With this geometry defined, the effect of meta-sheet stiffness can be examined from the force--displacement response required for impact mitigation. A simple mass-spring-damper model was used to compare three ideal restoring responses: a negative stiffness response (NSR), a zero stiffness response (ZSR), and a positive stiffness response (PSR) (Figure~2a). To evaluate the intrinsic mechanics of the stiffness profile, these responses were prescribed as simple cubic constitutive relations under nondimensional reference scales. The displacement, force, and stiffness scales were set to \(x_0=35.0\)~mm, \(F_0=1.8113\)~N, and \(K_0=F_0/x_0=51.75\)~N~m\(^{-1}\), respectively. The NSR curve has a normalized threshold peak force \(\bar{F}_{\mathrm{peak}}=1.00\) at \(\bar{x}=0.35\) and a local valley at \(\bar{x}=0.75\), defining a distinct negative stiffness region with \(\bar{k}<0\). The ZSR curve contains a quasi-zero-stiffness plateau with \(\bar{k}=0\) at \(\bar{x}=0.50\), while the PSR curve maintains monotonic elastic stiffening with \(\bar{k}>0\) throughout deformation. The prescribed force--displacement curves and their corresponding tangent stiffness profiles visualize these three cases (Figure~2b,c). Thus, the three ideal curves isolate how the sign and magnitude of the tangent stiffness affect impact mitigation before the specific kirigami geometry is tested experimentally. In normalized form, the local stiffness associated with the spring element in Figure~2a is
\begin{equation}
 \bar{k}(\bar{x})=\frac{\partial \bar{F}_{k}}{\partial \bar{x}}=\frac{\partial^2 \bar{U}}{\partial \bar{x}^2},
\label{eq:negative_stiffness}
\end{equation}
so that
\begin{align}
 &(NSR)\qquad
 \bar{k}>0\quad \mathrm{in~region~I},\qquad
 \bar{k}<0\quad \mathrm{in~region~II},\qquad
 \bar{k}>0\quad \mathrm{in~region~III}
 \\
 &(ZSR)\qquad
 \bar{k}>0\quad \mathrm{in~region~I},\qquad
 \bar{k}=0\quad \mathrm{in~region~II},\qquad
 \bar{k}>0\quad \mathrm{in~region~III}
 \\
 &(PSR)\qquad
 \bar{k}>0\quad \mathrm{in~all~regions}
\label{eq:stiffness_regions}
\end{align}

The dynamic consequence of these stiffness profiles was compared under an identical triangular compressive load history, which was linearly ramped to the maximum compressive load and then linearly returned to zero. The loading branch reached \(Q_{\max}=1.5F_0\) over \(T_{\mathrm{load}}=1.0\)~s. To isolate the contribution of the nonlinear stiffness profile, all extrinsic dynamic parameters were kept identical across the three cases, with \(m=0.01\)~kg and \(\xi=1.0\)~N~s~m\(^{-1}\) before nondimensionalization. Using the notation in Figure~2a, the normalized equation of motion is
\begin{equation}
 \bar{m}\ddot{\bar{x}}(\bar{t})+\bar{\xi}\dot{\bar{x}}(\bar{t})+\bar{F}_{k}(\bar{x})=\bar{Q}(\bar{t}),
\label{eq:eom}
\end{equation}
where \(\bar{m}\), \(\bar{\xi}\), \(\bar{F}_{k}\), \(\bar{Q}\), and \(\bar{x}\) correspond to the mass, damping coefficient, nonlinear restoring force of the stiffness element, applied load, and displacement, respectively. The corresponding viscous dissipation is
\begin{equation}
 \bar{E}_{\mathrm{visc}}(\bar{t})=
 \int_0^{\bar{t}}\bar{\xi}\dot{\bar{x}}(\tau)^2\,d\tau .
\label{eq:diss_power}
\end{equation}

Under the same triangular pulse with \(\bar{Q}_{\max}=1.5\), which is 1.5 times the normalized peak force, the PSR case resists deformation early and therefore uses only a limited structural stroke. The ZSR case increases stroke by introducing a force plateau, but the motion remains less energetic because the force does not drop into a softening region. In contrast, the NSR case enters a negative stiffness transition near \(\bar{Q}=1.0\) during loading and follows the corresponding unloading path, forming a dynamic hysteresis loop (Figure~2d). The associated increase in velocity during the negative stiffness transition generates sharp power dissipation peaks, consistent with the viscous power relation \(\bar{P}=\bar{\xi}\dot{\bar{x}}^2\). As a result, NSR produces the largest normalized dissipated energy, \(\bar{W}=0.46\), compared to \(\bar{W}=0.13\) for ZSR and \(\bar{W}=0.02\) for PSR (Figure~2e; Table~1). Note that the improvement is not simply a consequence of making the structure softer. It arises from the sequence of positive stiffness, negative stiffness, and hardening. The negative stiffness region reduces the transmitted force during deformation, while the hardening region limits excessive displacement and allows recovery of load capacity. This theoretical comparison provides an important insight into the kirigami meta-sheet design. Impact absorption is improved when a structure can enter a negative stiffness region and then return to a hardening region without sacrificial damage. Although buckling based absorbers have already adopted this principle, most of them arrange unit cells as vertical stacks or compression columns [19--28], limiting their practical use.

\subsection{Quasistatic Indentation for 3D-printed Kirigami Meta-sheets}
We first printed three kirigami meta-sheets with measured HR values of 0.08045 (LHR), 0.09739 (MHR), and 0.15682 (HHR), corresponding to mean measured hinge thicknesses of 1.1464, 1.3878, and 2.2347~mm, respectively (Figure~1e; Figure~S1; Table~S1; Table~S2). Quasistatic indentation experiments were then performed for these meta-sheets constrained by a reinforced polylactic acid (PLA) frame to characterize their force--displacement response (Figure~3a,b; Movie~S1). Detailed information on specimen preparation, indentation setup, and FE stress simulation setup is provided in the Experimental Section and Supporting Information (Table~S1; Supplementary Note~2; Figure~S2a,b; Figure~S3a,c; Figure~S4a).

The indentation response changes systematically with the hinge stiffness (Figure~3b,c; Table~2). The maximum depth of indentation for each experiment was selected as the largest deformation that could be applied without damaging the tied boundary and overloading the fishing-line fixation in the reinforced frame, resulting in approximately 35, 30, and 25~mm for LHR, MHR, and HHR, respectively (Figure~S2b). The LHR meta-sheet reaches the largest indentation depth and exhibits a local force maximum followed by a force reduction as a clear signature of the transition to a negative stiffness region. The HHR meta-sheet shows a monotonic stiffening response and approaches the PSR limit quickly. The MHR meta-sheet represents an intermediate case even though the zero-stiffness plateau is not as distinct as in the simple 1D model. As expected, increasing the hinge stiffness enhances the force carrying capacity, but suppresses the transition to the zero-to-negative stiffness.

The deformed shapes show hinge localized bending and lateral unit rotation, consistent with the intended mechanism of planar kirigami meta-sheets (Figure~3d). FE simulations performed for 25-mm indentation depth (Supplementary Note~2; Table~S2; Table~S3; Figure~S2a,b; Figure~S3a--d) following the previously reported analysis procedure [43] predict stresses concentrated near the hinge regions and hinge--unit junctions on the chuck side, confirming that the hinges are the structural elements that mediate entry into the negative stiffness region. The maximum von Mises stresses at the maximum indentation depth are 47.49, 40.60, and 32.95~MPa for LHR, MHR, and HHR, respectively. This stress localization would be a main factor to be considered in the impact absorber design, since the durability of kirigami meta-sheets depends on it.

The indentation hysteresis quantifies the recoverable energy loss pathway of the meta-sheet (Table~2). The loading work, unloading work, and dissipated energy are calculated as
\begin{equation}
W_{\mathrm{load}}=\int_{0}^{D_{\max}}F_{\mathrm{load}}(D)\,dD,
\quad
W_{\mathrm{unload}}=\int_{D_{\max}}^{0}F_{\mathrm{unload}}(D)\,dD,
\label{eq:work}
\end{equation}
\begin{equation}
E_{\mathrm{hys}}=W_{\mathrm{load}}-W_{\mathrm{unload}},
\quad
\eta=\frac{E_{\mathrm{hys}}}{W_{\mathrm{load}}}.
\label{eq:hys}
\end{equation}
Although the absolute dissipated energy of HHR is large because it carries a higher force, LHR provides the largest normalized dissipation and the most pronounced negative stiffness region. This distinction is important for impact absorption. A high force capacity alone does not guarantee efficient mitigation if the force is transmitted directly to the protected object. The LHR design instead sacrifices stiffness capacity to access a larger deformation stroke and higher normalized hysteresis.

\subsection{Drop-tower Impact Test and Dynamic Dissipation}
To investigate the effect of quasistatic stiffness on dynamic rebound and evaluate the impact absorption performance of kirigami meta-sheets, we performed drop-tower impact tests (Figure~4a--c; Movie~S2). The same 30~mm contact geometry and horizontally constrained specimens were used with velocity extraction and time calibration steps described in the Experimental Section and Supporting Information (Figure~S3b,d; Figure~S4b; Supplementary Notes~6 and~7; Table~S4). Surface acceleration was measured using an accelerometer attached one unit cell away from the impact point to capture transmitted vibration without occupying the direct contact region (Figure~4a; Figure~S4b). The measured impact velocities are nearly identical for the three samples (Table~3).

The rebound results follow the predicted hinge stiffness trend (Figure~4b; Table~3). During free fall, the three specimens show nearly identical slopes because the same gravitational acceleration acts on the chuck. After contact, the magnitude of the slope reflects impact deceleration (Figure~4b inset). This result indicates that the initial impact force increases in the order LHR \(<\) MHR \(<\) HHR, consistent with the decreased accessibility to the negative stiffness region as HR increases. High speed snapshots provide the same interpretation in physical space (Figure~4c). LHR undergoes the largest recoverable deformation before rebound, while HHR rebounds after a smaller deformation stroke.

We calculated the absorbed energy fraction for each trial, which is the ratio of rebound kinetic energy to incident kinetic energy, as
\begin{equation}
\phi_{\mathrm{abs}}=1-\frac{\frac{1}{2}Mv_r^2}{\frac{1}{2}Mv_0^2}=1-\left(\frac{v_r}{v_0}\right)^2=1-e^2
\label{eq:absorbed_fraction}
\end{equation}
where $e=\left|v_r/v_0\right|$ is the coefficient of restitution (Table~3). The specimen that most readily enters the buckling-induced negative stiffness region returns the least kinetic energy to the impactor, giving the highest absorbed energy fraction. LHR gives \(\phi_{\mathrm{abs}}=0.584\pm0.003\) larger than \(0.527\pm0.006\) and \(0.444\pm0.004\) corresponding to MHR and HHR, respectively. The equivalent damping coefficient follows the same order: \(c_{\mathrm{eq}}=0.9709\)~N\,s\,m\(^{-1}\) for LHR, \(0.7272\)~N\,s\,m\(^{-1}\) for MHR, and \(0.5645\)~N\,s\,m\(^{-1}\) for HHR. Thus, LHR has approximately 1.3 and 1.7 times the damping coefficient of MHR and HHR, respectively, indicating that the reduced rebound is accompanied by stronger velocity dependent dissipation. Thus, the negative stiffness region effectively increases the fraction of incident kinetic energy removed from rebound motion.

The force and dissipation measurements also provide a consistent dynamic interpretation (Figure~5a,b; Table~4). The calculated first impact force maintains the same monotonic order, LHR \(<\) MHR \(<\) HHR, confirming that compliant hinges soften the first collision. This ordering is especially important because the primary function of an impact absorber is to reduce the maximum force transmitted during the earliest contact event, when damage is most likely to occur. Representative post-processed acceleration-time traces are shown in the inset of Figure~5b. Using the equivalent damping coefficients in Table~3, the dissipated energy based on the accelerometer was estimated as
\begin{equation}
E_{\mathrm{acc}}(t)=c_{\mathrm{eq}}\int_0^t v_{\mathrm{acc}}(\tau)^2\,d\tau,
\label{eq:Eacc}
\end{equation}
where \(v_{\mathrm{acc}}\) is the local vibration velocity reconstructed from the accelerometer signal. This expression follows the viscous dissipation form used in the simple dynamic model: larger \(c_{\mathrm{eq}}\) and larger vibration velocity both increase dissipated energy. The sensor was placed one unit cell away from the contact point, so \(E_{\mathrm{acc}}\) is used as a reference indicator rather than a full field energy balance. This metric nevertheless gives \(10.36\pm0.70\)~mJ for LHR, which is about 4.7 and 6.2 times the MHR and HHR values, respectively. These consistent results support that entering the negative stiffness region reduces rebound, lowers the first impact force, and increases vibration mediated energy dissipation.

\subsection{Egg Drop Test}
To demonstrate the protection capability of kirigami meta-sheets, we performed egg drop experiments using a fixed release setup (Figure~6a; Movie~S3). A larger kirigami meta-sheet, a 3.2~mm thick polyethylene (PE) mesh, and a 10~mm thick styrofoam panel were tested under the same release condition (Figure~6b--d). The projected area of the kirigami meta-sheet used for this demonstration was four times larger than the one used in the drop-tower impact test. We used jumbo eggs whose mass was greater than 68~g. A 1.5~m height release clamp assembly was fabricated to place an egg at the same release position for every trial. The absorber surface was 73~mm above the ground and the free drop height from the egg to the absorber surface was 1383.8~mm.

The kirigami meta-sheet is the only absorber that protects both the egg and the absorber (Figure~6b). After impact, the egg remains intact, and the meta-sheet does not show any visible tearing, hinge fracture, or permanent crushing. In contrast, the PE mesh was too stiff to absorb the impact energy and prevent the egg from being broken (Figure~6c). The styrofoam representing a conventional sacrificial absorber fractured during the impact and could not protect the egg (Figure~6d). This result confirms that the proposed kirigami meta-sheet may offer an effective way of programming impact absorption capability in thin structures.

\section{Conclusion}
A planar kirigami meta-sheet is proposed as an impact absorber that translates local out-of-plane impact into distributed hinge buckling, unit rotation, stiffness regime transition, and vibration. Negative stiffness improves impact mitigation by allowing the structure to enter a softening region and then return to a hardening region without sacrificial crushing. The kirigami meta-sheet realizes this response in a planar architecture, overcoming the limitation of vertical stacking and backing surface in many metamaterial-based absorbers. Drop-tower experiments clearly show that the kirigami meta-sheet with a positive-negative-positive stiffness transition exhibits the lowest rebound coefficient, the highest absorbed energy fraction, the smallest first impact force, and the largest dissipation.

Together, kirigami meta-sheets with negative stiffness transitions offer practical planar architectures for scalable impact attenuation. This planar architecture can be easily enlarged by geometric scaling, tiled over broad areas, and manufactured by scalable sheet based processes such as laser cutting, casting, molding, or lamination. For repeated use of kirigami meta-sheets, further studies should focus on fatigue resistant hinge design, multilayer meta-sheet assemblies, standardized peak force measurements, and scalable manufacturing. In particular, hinge reinforcement and elastomeric lamination are promising routes to suppress hinge stress concentration while preserving the negative stiffness transition that gives the meta-sheet its energy absorbing advantage.

\section{Experimental Section}
\threesubsection{Specimen Preparation}\par
Meta-sheet specimens were fabricated as kirigami specimens designed to deploy into a dome-shaped configuration using the validated Ultimaker S5 printing protocol (Table~S1). The nominal HR values were designed as 0.06250, 0.09000, and 0.15000 for the LHR, MHR, and HHR configurations, respectively, and the designed patterns were printed as flat sheets (Figure~S1a). The as fabricated hinge thicknesses were then measured optically because the narrow hinges can deviate from the nominal design during printing; the measured thicknesses were used to calculate calibrated HR values of 0.08045, 0.09739, and 0.15682 (Figure~S1b; Table~S2). The specimens were fixed tightly to reinforced PLA frames using fishing lines and adhesive, reproducing the horizontal boundary constraint used in the finite element model (Figure~S2a,b). This dome-forming kirigami geometry allowed a repeated unit response to be measured while maintaining compatibility with the indentation and drop-tower fixtures.

\threesubsection{Simple Dynamic Simulation}\par
The simple mass-spring-damper simulations in Figure~2 were solved using numerical time integration of Equation~\ref{eq:eom}. Initial conditions were $\bar{x}(0)=0$ and $\dot{\bar{x}}(0)=0$, and the simulations were solved with a maximum integration step of $5\times10^{-5}$. Dissipation power and cumulative normalized viscous energy were calculated from Equation~\ref{eq:diss_power}.

\threesubsection{Quasistatic Indentation}\par
Quasistatic indentation was performed at a crosshead speed of 80~mm~min$^{-1}$, corresponding to 1.3333~mm~s$^{-1}$, using the 30~mm diameter indentation chuck and the leveled two-lift setup (Figure~S3a,c; Figure~S4a). The indentation experiments are provided in the Supporting Information in the order LHR, MHR, and HHR; for each specimen group, the front-view indentation recording is followed by the corresponding bottom-view recording (Movie~S1). Hysteresis energy was calculated by numerical integration of the measured loading and unloading force--displacement curves.

\threesubsection{Material Properties and ABAQUS Stress Simulation}\par
The physical parameters used in experiments and simulations were determined from direct measurements whenever possible. The calibrated hinge geometry was used for each FE model. Specimen density was calculated from measured mass divided by the computer aided design (CAD) volume of the corresponding kirigami array. The elastic modulus used in ABAQUS was obtained from tensile calibration of printed strips by fitting the initial linear stress--strain response; the Poisson's ratio was taken from the calibrated material model used for all simulations. The FE quasistatic indentation model used the calibrated hinge geometry, calibrated elastic properties, experimental indenter geometry, and the same edge constraints as the physical setup. The simulations were used to inspect hinge buckling, dome inversion, lateral spreading, and hinge localized stress in the LHR, MHR, and HHR specimens.

\threesubsection{Drop-tower Impact and High Speed Analysis}\par
Drop-tower impact experiments were performed using a measured drop assembly mass of 0.203~kg and the 30~mm diameter drop chuck (Figure~S3b,d). The synchronized impact setup combined electromagnetic release, high speed imaging, and accelerometer recording (Figure~S4b). The impact and rebound velocities were extracted from CHRONOS 1.4 high speed camera height data, and additional SONY RX10 III recordings were used to visualize kirigami deformation during indentation and impact (Movie~S2). The free fall portion of the trajectory was fitted using a quadratic height--time relation to obtain a camera time correction factor $K=1.4359$,
\begin{equation}
h(t_{\mathrm{meas}})=h_0-\frac{1}{2}gK^2t_{\mathrm{meas}}^2
=h_0-\frac{1}{2}a_{\mathrm{meas}}t_{\mathrm{meas}}^2,
\qquad
K=\sqrt{\frac{a_{\mathrm{meas}}}{g}},
\label{eq:Kfactor}
\end{equation}
and the physical time and image derived velocity were calculated as
\begin{equation}
t_{\mathrm{real}}=Kt_{\mathrm{meas}},\qquad
v(t)=\frac{dh}{dt_{\mathrm{real}}}.
\label{eq:video_velocity}
\end{equation}
The calibrated mean impact velocities are reported in the Supporting Information (Table~S4). Sensor data from the accelerometer and force sensor were recorded using the hardware time base and were therefore not corrected by this camera factor.

\threesubsection{Accelerometer and Force Sensor Processing}\par
The accelerometer was positioned one unit cell away from the impact point, at the middle of the boomerang-shaped connector between neighboring triangular units, to measure transmitted vibration without occupying the contact region (Supplementary Note~5; Figure~S4b). The accelerometer sensitivity was 10~mV~(m~s$^{-2}$)$^{-1}$ and the force sensor sensitivity was 100~mV~N$^{-1}$. The voltage to acceleration conversion, baseline correction, high pass filtering, and time alignment are given in the Supporting Information (Supplementary Notes~8 and~9). After offset removal, acceleration and force were converted from voltage signals as
\begin{equation}
a(t)=\frac{V_{\mathrm{acc}}(t)-V_{\mathrm{acc},0}}{0.01},\qquad
F_{\mathrm{sensor}}(t)=\frac{V_F(t)-V_{F,0}}{0.1},
\label{eq:sensor_conversion}
\end{equation}
where 0.01~V~(m~s$^{-2}$)$^{-1}$ and 0.1~V~N$^{-1}$ are the accelerometer and force sensor sensitivities, respectively. Acceleration data were offset corrected and high pass filtered using a 2~Hz Butterworth filter before integration. Velocity and displacement were calculated by numerical integration,
\begin{equation}
v(t)=v(0)+\int_0^t a(\tau)\,d\tau,\qquad
x(t)=x(0)+\int_0^t v(\tau)\,d\tau .
\label{eq:accel_integration}
\end{equation}
Impact start time was identified by tracing backward from the first force value exceeding 3~N to the nearest point below 1~N, and impact end time was identified after the peak when force first decreased below 1~N. For the approximately 400~mm free-fall drop, the theoretical impact velocity is about 2.8~m~s$^{-1}$, so the physically expected impact-induced velocity change is on the order of 2.8--5.6~m~s$^{-1}$ between the perfectly inelastic and ideal elastic rebound limits. A conservative 1--8~m~s$^{-1}$ acceptance range was therefore used to tolerate experimental noise, baseline drift, and small off-axis effects while excluding trials dominated by sensor artifacts or trigger failures. Trials with velocity changes outside this range were excluded from damping analysis. Equivalent damping was calculated as
\begin{equation}
\xi=c_{\mathrm{eq}}=
\frac{\displaystyle\int_{t_s}^{t_e}F_{\mathrm{sensor}}(t)v(t)\,dt}
{\displaystyle\int_{t_s}^{t_e}v(t)^2\,dt},
\label{eq:ceq}
\end{equation}
where $t_s$ and $t_e$ are the impact start and end times. The chuck velocity used in Equation~\ref{eq:ceq} was reconstructed from Newton's second law as
\begin{equation}
v(t)=v_0-\int_{t_s}^{t}\frac{F_{\mathrm{sensor}}(\tau)}{M_{\mathrm{chuck}}}\,d\tau,
\label{eq:newton_velocity}
\end{equation}
with $M_{\mathrm{chuck}}=0.203$~kg.

\threesubsection{Egg Drop Test}\par
Egg drop experiments used jumbo eggs with mass greater than 68~g, a larger kirigami meta-sheet, a 3.2~mm thick PE mesh, and 10~mm thick styrofoam. A 1.5~m height release clamp assembly was fabricated to keep the egg position identical across tests. The release height measured from the ground was 1456.8~mm, and the absorber surface was 73~mm above the ground, giving a free drop height of 1383.8~mm from the egg to the absorber surface (Movie~S3).

\raggedright

\medskip
\textbf{Supporting Information} \par
% Publisher-specific Supporting Information availability notice removed for arXiv.

\medskip
\textbf{Acknowledgements} \par
This work was supported by the National Research Foundation of Korea(NRF) grant funded by the Korea government(MSIT and MOE) (No. RS-2026-25500579).

\medskip
\textbf{Conflict of Interest} \par
The authors declare no conflict of interest.

\medskip
\textbf{Data Availability Statement} \par
The data that support the plots within this paper and other findings of this study are available from the corresponding author upon reasonable request.

\medskip
\textbf{References}\\
\noindent 1. P. Qiao, M. Yang, F. Bobaru, ``Impact Mechanics and High-Energy Absorbing Materials: Review,'' \textit{Journal of Aerospace Engineering} \textbf{2008}, \textit{21}, 235--248. https://doi.org/10.1061/(ASCE)0893-1321(2008)21:4(235)\\
2. A. A. A. Alghamdi, ``Collapsible Impact Energy Absorbers: An Overview,'' \textit{Thin-Walled Structures} \textbf{2001}, \textit{39}, 189--213. https://doi.org/10.1016/S0263-8231(00)00048-3\\
3. S. R. Reid, ``Plastic Deformation Mechanisms in Axially Compressed Metal Tubes Used as Impact Energy Absorbers,'' \textit{International Journal of Mechanical Sciences} \textbf{1993}, \textit{35}, 1035--1052. https://doi.org/10.1016/0020-7403(93)90054-X\\
4. H. G. Tattersall, G. Tappin, ``The Work of Fracture and Its Measurement in Metals, Ceramics and Other Materials,'' \textit{Journal of Materials Science} \textbf{1966}, \textit{1}, 296--301. https://doi.org/10.1007/BF00550177\\
5. M. A. Dawson, G. H. McKinley, L. J. Gibson, ``The Dynamic Compressive Response of an Open-Cell Foam Impregnated With a Non-Newtonian Fluid,'' \textit{Journal of Applied Mechanics} \textbf{2009}, \textit{76}, 061011. https://doi.org/10.1115/1.3130825\\
6. M. Avalle, G. Belingardi, R. Montanini, ``Characterization of Polymeric Structural Foams under Compressive Impact Loading by Means of Energy-Absorption Diagram,'' \textit{International Journal of Impact Engineering} \textbf{2001}, \textit{25}, 455--472. https://doi.org/10.1016/S0734-743X(00)00060-9\\
7. V. S. Deshpande, N. A. Fleck, ``High Strain Rate Compressive Behaviour of Aluminium Alloy Foams,'' \textit{International Journal of Impact Engineering} \textbf{2000}, \textit{24}, 277--298. https://doi.org/10.1016/S0734-743X(99)00153-0\\
8. K. Fu, Z. Zhao, L. Jin, ``Programmable Granular Metamaterials for Reusable Energy Absorption,'' \textit{Advanced Functional Materials} \textbf{2019}, \textit{29}, 1901258. https://doi.org/10.1002/adfm.201901258\\
9. T. A. Schaedler, A. J. Jacobsen, A. Torrents, A. E. Sorensen, J. Lian, J. R. Greer, L. Valdevit, W. B. Carter, ``Ultralight Metallic Microlattices,'' \textit{Science} \textbf{2011}, \textit{334}, 962--965. https://doi.org/10.1126/science.1211649\\
10. L. R. Meza, S. Das, J. R. Greer, ``Strong, Lightweight, and Recoverable Three-Dimensional Ceramic Nanolattices,'' \textit{Science} \textbf{2014}, \textit{345}, 1322--1326. https://doi.org/10.1126/science.1255908\\
11. X. Zheng, H. Lee, T. H. Weisgraber, M. Shusteff, J. DeOtte, et al., ``Ultralight, Ultrastiff Mechanical Metamaterials,'' \textit{Science} \textbf{2014}, \textit{344}, 1373--1377. https://doi.org/10.1126/science.1252291\\
12. K. Bertoldi, V. Vitelli, J. Christensen, M. van Hecke, ``Flexible Mechanical Metamaterials,'' \textit{Nature Reviews Materials} \textbf{2017}, \textit{2}, 17066. https://doi.org/10.1038/natrevmats.2017.66\\
13. S. Babaee, J. Shim, J. C. Weaver, E. R. Chen, N. Patel, K. Bertoldi, ``3D Soft Metamaterials with Negative Poisson's Ratio,'' \textit{Advanced Materials} \textbf{2013}, \textit{25}, 5044--5049. https://doi.org/10.1002/adma.201301986\\
14. J. N. Grima, A. Alderson, K. E. Evans, ``Auxetic Behaviour from Rotating Rigid Units,'' \textit{physica status solidi (b)} \textbf{2005}, \textit{242}, 561--575. https://doi.org/10.1002/pssb.200460376\\
15. X. J. Tan, B. Wang, S. W. Zhu, S. Chen, K. Yao, P. F. Xu, L. Z. Wu, Y. G. Sun, ``Novel Multidirectional Negative Stiffness Mechanical Metamaterials,'' \textit{Smart Materials and Structures} \textbf{2020}, \textit{29}, 015037. https://doi.org/10.1088/1361-665X/ab47d9\\
16. B. Florijn, C. Coulais, M. van Hecke, ``Programmable Mechanical Metamaterials,'' \textit{Physical Review Letters} \textbf{2014}, \textit{113}, 175503. https://doi.org/10.1103/PhysRevLett.113.175503\\
17. C. Coulais, D. Sounas, A. Alu, ``Static Non-Reciprocity in Mechanical Metamaterials,'' \textit{Nature} \textbf{2017}, \textit{542}, 461--464. https://doi.org/10.1038/nature21044\\
18. B. Haghpanah, L. Salari-Sharif, P. Pourrajab, J. Hopkins, L. Valdevit, ``Multistable Shape-Reconfigurable Architected Materials,'' \textit{Advanced Materials} \textbf{2016}, \textit{28}, 7915--7920. https://doi.org/10.1002/adma.201601650\\
19. T. Frenzel, C. Findeisen, M. Kadic, P. Gumbsch, M. Wegener, ``Tailored Buckling Microlattices as Reusable Light-Weight Shock Absorbers,'' \textit{Advanced Materials} \textbf{2016}, \textit{28}, 5865--5870. https://doi.org/10.1002/adma.201600610\\
20. D. Restrepo, N. D. Mankame, P. D. Zavattieri, ``Phase Transforming Cellular Materials,'' \textit{Extreme Mechanics Letters} \textbf{2015}, \textit{4}, 52--60. https://doi.org/10.1016/j.eml.2015.08.001\\
21. S. Shan, S. H. Kang, J. R. Raney, P. Wang, L. Fang, F. Candido, J. A. Lewis, K. Bertoldi, ``Multistable Architected Materials for Trapping Elastic Strain Energy,'' \textit{Advanced Materials} \textbf{2015}, \textit{27}, 4296--4301. https://doi.org/10.1002/adma.201501708\\
22. J. R. Raney, N. Nadkarni, C. Daraio, D. M. Kochmann, J. A. Lewis, K. Bertoldi, ``Stable Propagation of Mechanical Signals in Soft Media Using Stored Elastic Energy,'' \textit{Proceedings of the National Academy of Sciences of the United States of America} \textbf{2016}, \textit{113}, 9722--9727. https://doi.org/10.1073/pnas.1604838113\\
23. D. M. Correa, T. Klatt, S. Cortes, M. Haberman, D. Kovar, C. Seepersad, ``Negative Stiffness Honeycombs for Recoverable Shock Isolation,'' \textit{Rapid Prototyping Journal} \textbf{2015}, \textit{21}, 193--200. https://doi.org/10.1108/RPJ-12-2014-0182\\
24. C. Findeisen, J. Hohe, M. Kadic, P. Gumbsch, ``Characteristics of Mechanical Metamaterials Based on Buckling Elements,'' \textit{Journal of the Mechanics and Physics of Solids} \textbf{2017}, \textit{102}, 151--164. https://doi.org/10.1016/j.jmps.2017.02.011\\
25. J. T. B. Overvelde, S. Shan, K. Bertoldi, ``Compaction Through Buckling in 2D Periodic, Soft and Porous Structures: Effect of Pore Shape,'' \textit{Advanced Materials} \textbf{2012}, \textit{24}, 2337--2342. https://doi.org/10.1002/adma.201104395\\
26. S. Ji, F. Wang, J. Wang, Z. Wang, C. Wang, Y. Wei, ``Dynamic Responses and Energy Absorption of Mechanical Metamaterials Composed of Buckling Beams,'' \textit{Journal of Vibration Engineering \& Technologies} \textbf{2024}, \textit{12}, 1249--1261. https://doi.org/10.1007/s42417-023-00904-w\\
27. S. Chen, X. Tan, J. Hu, B. Wang, L. Wang, Y. Zou, L. Wu, ``Continuous Carbon Fiber Reinforced Composite Negative Stiffness Mechanical Metamaterial for Recoverable Energy Absorption,'' \textit{Composite Structures} \textbf{2022}, \textit{288}, 115411. https://doi.org/10.1016/j.compstruct.2022.115411\\
28. R. A. Ibrahim, ``Recent Advances in Nonlinear Passive Vibration Isolators,'' \textit{Journal of Sound and Vibration} \textbf{2008}, \textit{314}, 371--452. https://doi.org/10.1016/j.jsv.2008.01.014\\
29. T. C. Shyu, P. F. Damasceno, P. M. Dodd, A. Lamoureux, L. Z. Xu, M. Shlian, M. Shtein, S. C. Glotzer, N. A. Kotov, ``A Kirigami Approach to Engineering Elasticity in Nanocomposites through Patterned Defects,'' \textit{Nature Materials} \textbf{2015}, \textit{14}, 785--789. https://doi.org/10.1038/nmat4327\\
30. S. J. P. Callens, A. A. Zadpoor, ``From Flat Sheets to Curved Geometries: Origami and Kirigami Approaches,'' \textit{Materials Today} \textbf{2018}, \textit{21}, 241--264. https://doi.org/10.1016/j.mattod.2017.10.004\\
31. T. Castle, Y. Cho, X. Gong, E. Jung, D. M. Sussman, S. Yang, R. D. Kamien, ``Making the Cut: Lattice Kirigami Rules,'' \textit{Physical Review Letters} \textbf{2014}, \textit{113}, 245502. https://doi.org/10.1103/PhysRevLett.113.245502\\
32. M. K. Blees, A. W. Barnard, P. A. Rose, S. P. Roberts, K. L. McGill, P. Y. Huang, A. R. Ruyack, J. W. Kevek, B. Kobrin, D. A. Muller, P. L. McEuen, ``Graphene Kirigami,'' \textit{Nature} \textbf{2015}, \textit{524}, 204--207. https://doi.org/10.1038/nature14588\\
33. D. M. Sussman, Y. Cho, T. Castle, X. Gong, E. Jung, S. Yang, R. D. Kamien, ``Algorithmic Lattice Kirigami: A Route to Pluripotent Materials,'' \textit{Proceedings of the National Academy of Sciences of the United States of America} \textbf{2015}, \textit{112}, 7449--7453. https://doi.org/10.1073/pnas.1506048112\\
34. T. Castle, D. M. Sussman, M. Tanis, R. D. Kamien, ``Additive Lattice Kirigami,'' \textit{Science Advances} \textbf{2016}, \textit{2}, e1601258. https://doi.org/10.1126/sciadv.1601258\\
35. M. Isobe, K. Okumura, ``Initial Rigid Response and Softening Transition of Highly Stretchable Kirigami Sheet Materials,'' \textit{Scientific Reports} \textbf{2016}, \textit{6}, 24758. https://doi.org/10.1038/srep24758\\
36. Y. Tang, J. Yin, ``Design of Cut Unit Geometry in Hierarchical Kirigami-Based Auxetic Metamaterials for High Stretchability and Compressibility,'' \textit{Extreme Mechanics Letters} \textbf{2017}, \textit{12}, 77--85. https://doi.org/10.1016/j.eml.2016.07.005\\
37. A. Rafsanjani, K. Bertoldi, ``Buckling-Induced Kirigami,'' \textit{Physical Review Letters} \textbf{2017}, \textit{118}, 084301. https://doi.org/10.1103/PhysRevLett.118.084301\\
38. Y. Tang, G. Lin, S. Yang, Y. K. Yi, R. D. Kamien, J. Yin, ``Programmable Kiri-Kirigami Metamaterials,'' \textit{Advanced Materials} \textbf{2017}, \textit{29}, 1604262. https://doi.org/10.1002/adma.201604262\\
39. A. Rafsanjani, L. Jin, B. Deng, K. Bertoldi, ``Propagation of Pop Ups in Kirigami Shells,'' \textit{Proceedings of the National Academy of Sciences of the United States of America} \textbf{2019}, \textit{116}, 8200--8205. https://doi.org/10.1073/pnas.1817763116\\
40. A. Rafsanjani, Y. Zhang, B. Liu, S. M. Rubinstein, K. Bertoldi, ``Kirigami Skins Make a Simple Soft Actuator Crawl,'' \textit{Science Robotics} \textbf{2018}, \textit{3}, eaar7555. https://doi.org/10.1126/scirobotics.aar7555\\
41. N. An, A. G. Domel, J. Zhou, A. Rafsanjani, K. Bertoldi, ``Programmable Hierarchical Kirigami,'' \textit{Advanced Functional Materials} \textbf{2020}, \textit{30}, 1906711. https://doi.org/10.1002/adfm.201906711\\
42. J. M. Hur, D.-N. Kim, ``Auxetic Meta-Disk for Independent Control of Flexural and Torsional Waves,'' \textit{International Journal of Mechanical Sciences} \textbf{2023}, \textit{243}, 108050. https://doi.org/10.1016/j.ijmecsci.2022.108050\\
43. H. Cho, D.-N. Kim, ``Controlling the Stiffness of Bistable Kirigami Surfaces via Spatially Varying Hinges,'' \textit{Materials \& Design} \textbf{2023}, \textit{231}, 112053. https://doi.org/10.1016/j.matdes.2023.112053\\
44. Y. Do, D.-N. Kim, ``Investigating the Effect of Slit Patterns on the Deformation of Thin-Walled Tubes under Side Impact,'' \textit{Journal of Mechanical Science and Technology} \textbf{2022}, \textit{36}, 5649--5655. https://doi.org/10.1007/s12206-022-1027-4\\

\clearpage

\begin{figure}[H]
  \includegraphics[width=\linewidth, trim={0.0cm 10.5cm 0.0cm 0.0cm}]{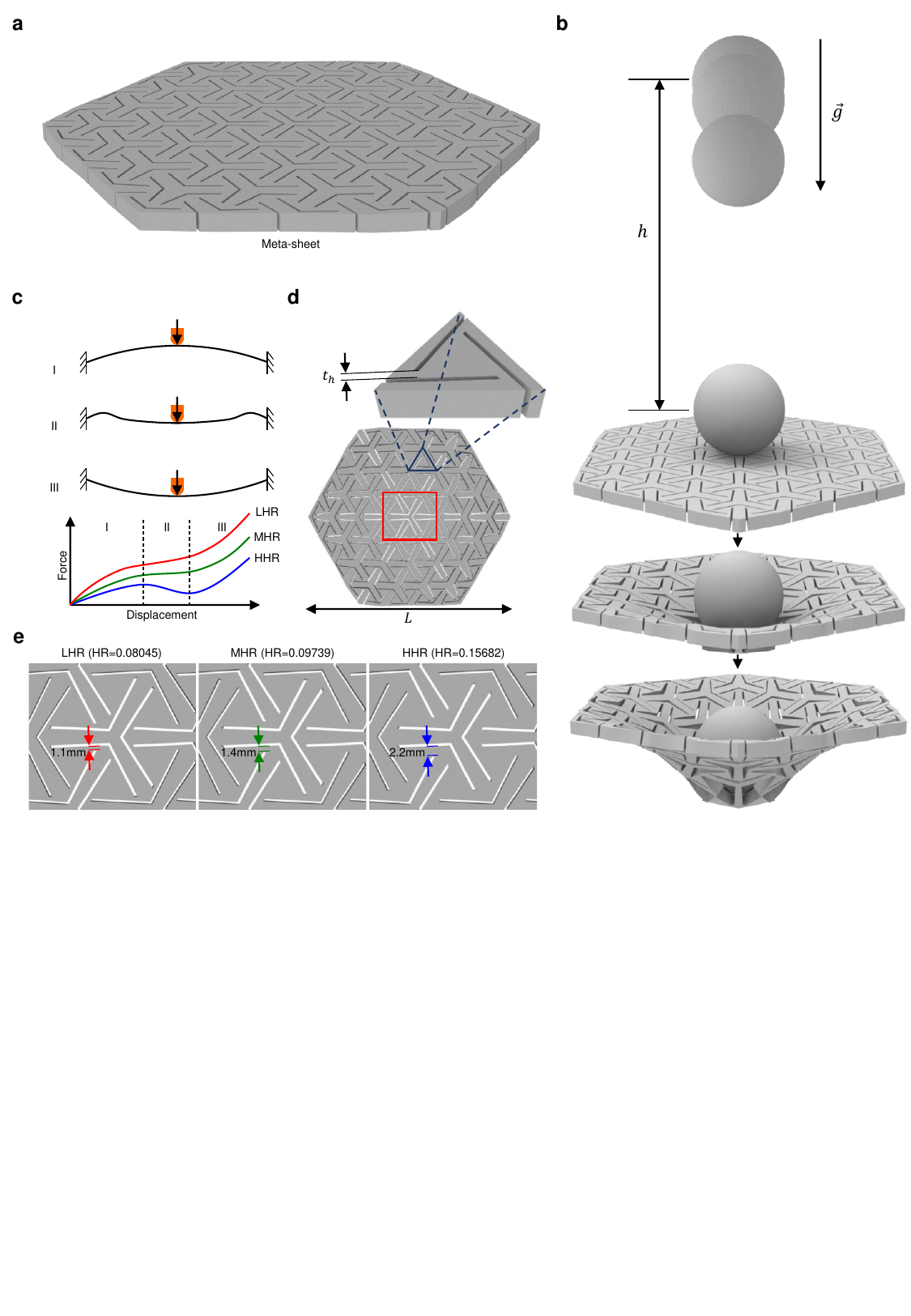}
  \caption{Kirigami meta-sheet design. (a) Meta-sheet geometry. (b) Drop impact schematic with release height $h$, gravitational acceleration $\vec{g}$, and deformation snapshots. (c) Buckling based force--displacement regions I, II, and III with representative LHR, MHR, and HHR responses. (d) Hinge thickness $t_h$, pattern length $L$, and magnified hinge region. (e) LHR, MHR, and HHR specimens with highlighted hinge regions, measured hinge thicknesses, and HR values.}
  \label{fig:fig1}
\end{figure}

\begin{figure}[H]
  \includegraphics[width=\linewidth, trim={0.0cm 8.7cm 0.0cm 0.0cm}]{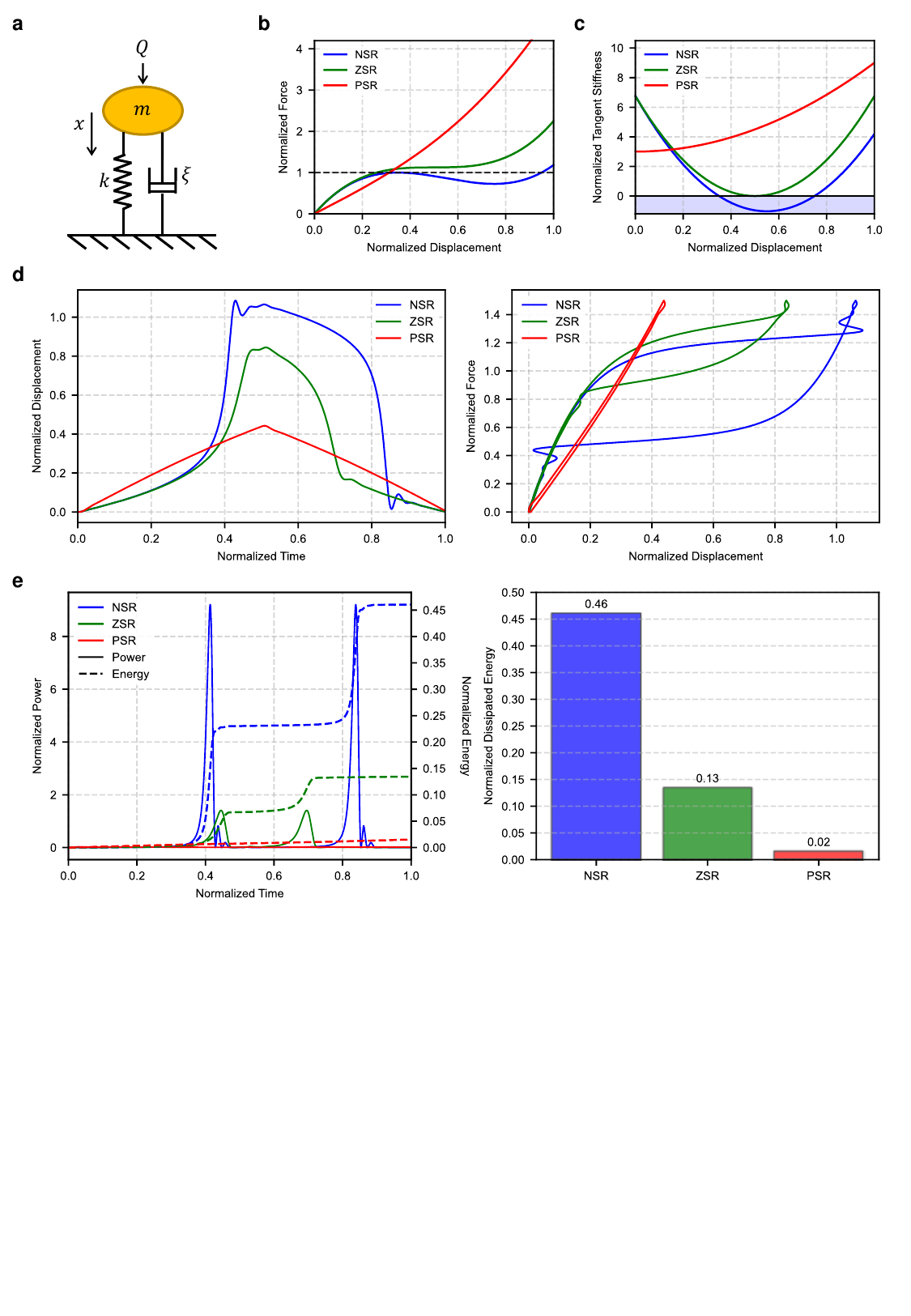}
  \caption{Simple dynamic model for stiffness profile selection. (a) One degree of freedom model. (b) Ideal nondimensional force--displacement curves. (c) Nondimensional stiffness. (d) Nondimensional displacement histories and force--displacement loops. (e) Nondimensional dissipation power and cumulative dissipated energy.}
  \label{fig:fig2}
\end{figure}

\begin{figure}[H]
  \includegraphics[width=\linewidth, trim={0.0cm 9.5cm 0.0cm 0.0cm}]{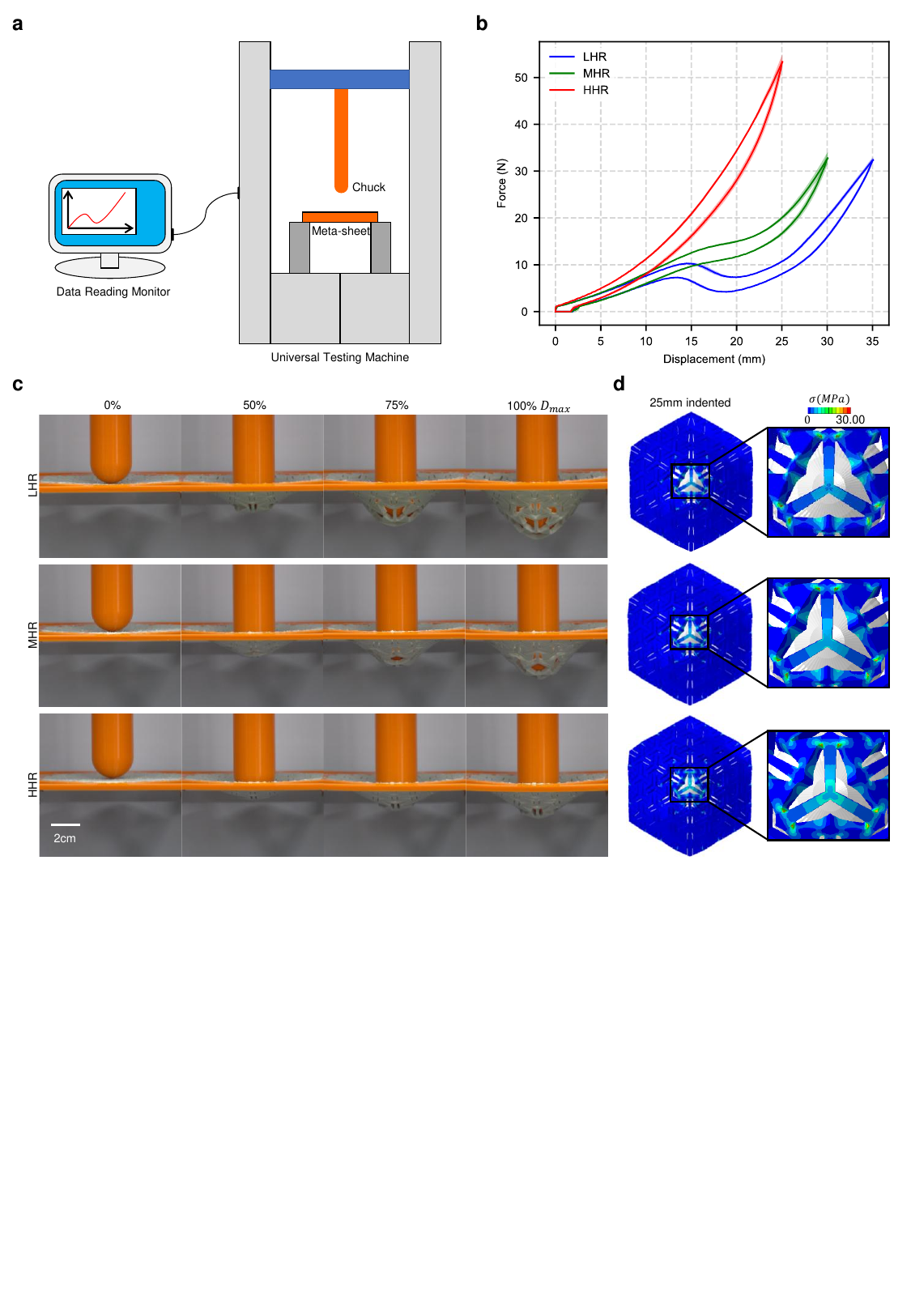}
  \caption{Quasistatic indentation and FE stress analysis. (a) Indentation setup. (b) Experimental force--displacement curves. (c) Indentation image sequence at 0\%, 50\%, 75\%, and 100\% of the specimen-specific $D_{\max}$. (d) FE von Mises stress contours at 25~mm indentation with a common 0--30~MPa color range.}
  \label{fig:fig3}
\end{figure}

\begin{figure}[H]
  \includegraphics[width=\linewidth, trim={0.0cm 6.3cm 0.0cm 0.0cm}]{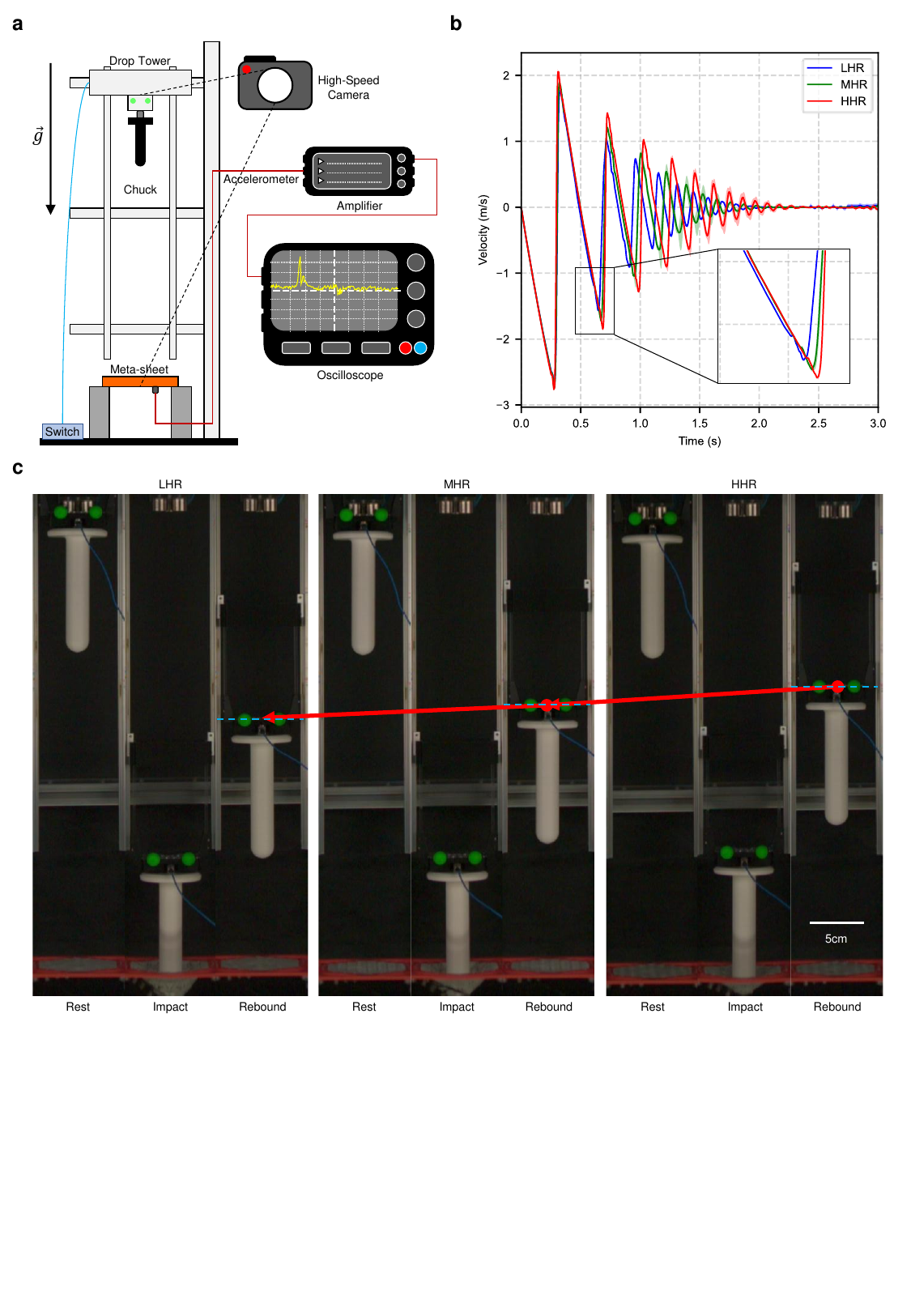}
  \caption{Drop-tower impact tests. (a) Experimental setup with gravitational acceleration direction, drop-tower, chuck, meta-sheet, high speed camera, accelerometer, amplifier, oscilloscope, and switch. (b) Velocity histories for LHR, MHR, and HHR specimens with a magnified inset near first contact. (c) High speed snapshots at rest, impact, and rebound.}
  \label{fig:fig4}
\end{figure}

\begin{figure}[H]
  \includegraphics[width=\linewidth, trim={0.0cm 18.5cm 0.0cm 0.0cm}]{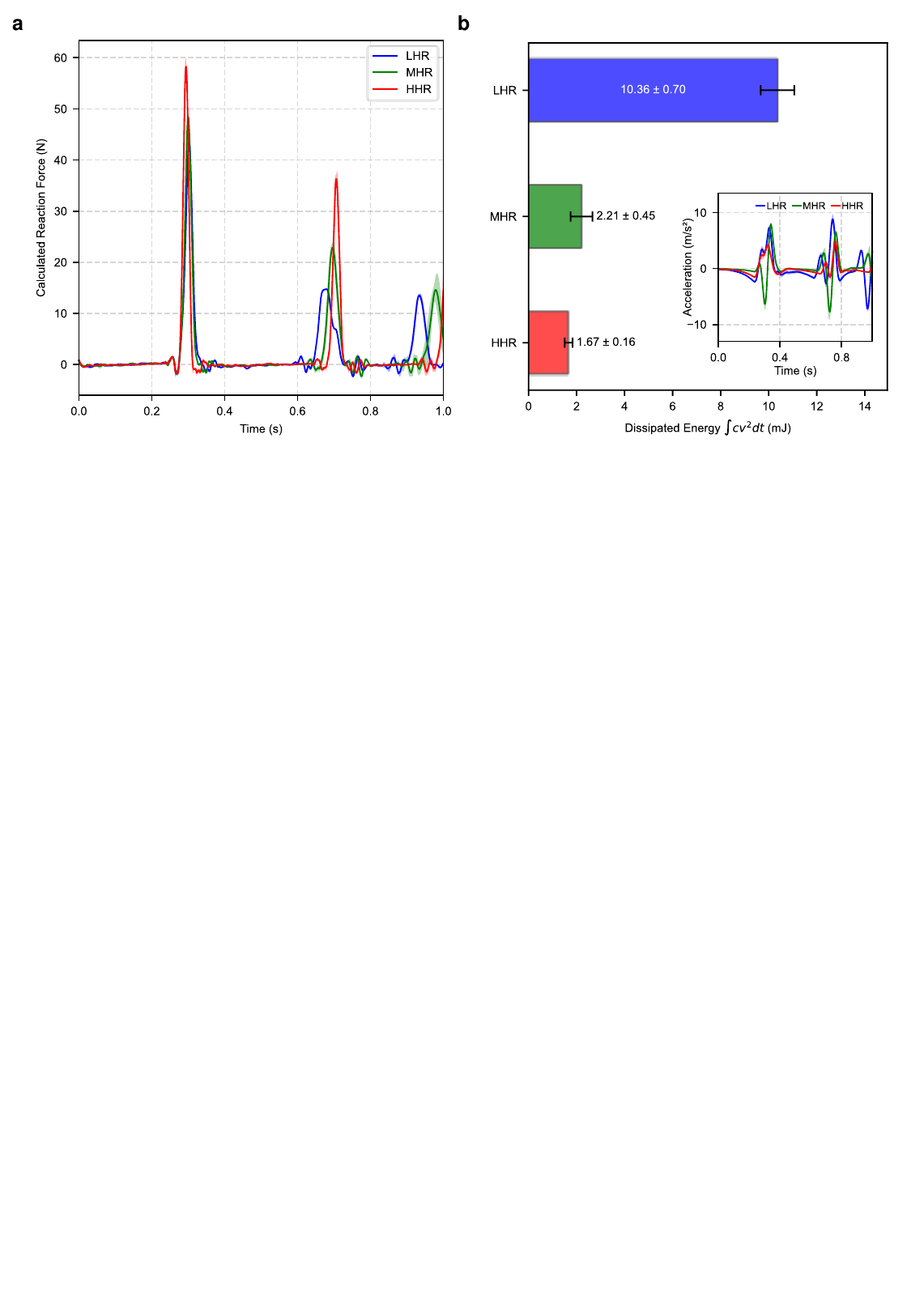}
  \caption{Drop impact force and dissipation metrics. (a) Calculated reaction force histories for LHR, MHR, and HHR specimens. (b) Dissipated energy based on the accelerometer with an inset showing representative post-processed acceleration-time traces.}
  \label{fig:fig5}
\end{figure}

\begin{figure}[H]
  \includegraphics[width=\linewidth, trim={0.0cm 10.0cm 0.0cm 0.0cm}]{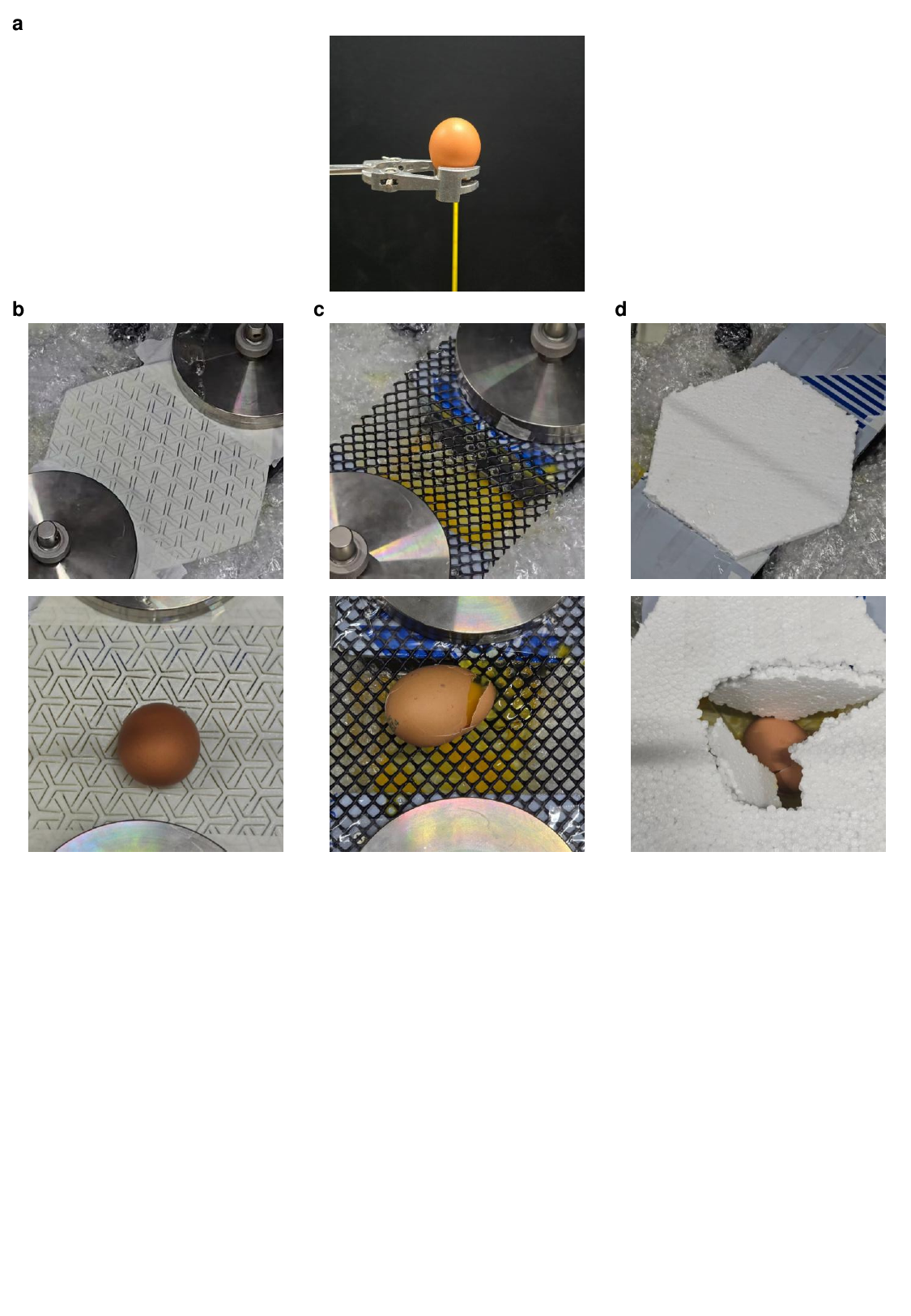}
  \caption{Egg drop test. (a) Egg drop setup. (b) Larger kirigami meta-sheet. (c) 3.2~mm thick PE mesh. (d) 10~mm thick styrofoam panel.}
  \label{fig:fig6}
\end{figure}

\begin{table}[H]
  \centering
  \caption{Simple dynamic model outputs for the three ideal stiffness responses.}
  \begin{tabular}{@{}llll@{}}
    \toprule
    Response & Max disp. & Max power & Total dissipated energy \\
    \midrule
    NSR & 1.09 & 9.21 & 0.46 \\
    ZSR & 0.84 & 1.41 & 0.13 \\
    PSR & 0.44 & 0.03 & 0.02 \\
    \bottomrule
  \end{tabular}
  \label{tab:normalized_model}
\end{table}

\begin{table}[H]
  \centering
  \caption{Quasistatic indentation metrics of dome-forming kirigami specimens.}
  \begin{tabular}{@{}llllll@{}}
    \toprule
    Specimen & Hinge thickness $t_h$ (mm) & HR & Max disp. (mm) & Max force (N) & Hysteresis ratio (\%) \\
    \midrule
    LHR & 1.1464 & 0.08045 & 35.07 & 32.379 & 24.28 \\
    MHR & 1.3878 & 0.09739 & 30.06 & 32.775 & 19.76 \\
    HHR & 2.2347 & 0.15682 & 25.07 & 53.351 & 19.34 \\
    \bottomrule
  \end{tabular}
  \label{tab:indentation}
\end{table}

\begin{table}[H]
  \centering
  \caption{Drop-tower rebound and damping metrics.}
  \begin{tabular}{@{}llllll@{}}
    \toprule
    Specimen & $v_0$ (m\,s$^{-1}$) & $v_r$ (m\,s$^{-1}$) & $e$ & $\phi_{\mathrm{abs}}$ & $c_{\mathrm{eq}}$ (N\,s\,m$^{-1}$) \\
    \midrule
    LHR & 2.7664 & 1.7746 & 0.6415 & 0.584 $\pm$ 0.003 & 0.9709 \\
    MHR & 2.7657 & 1.8977 & 0.6861 & 0.527 $\pm$ 0.006 & 0.7272 \\
    HHR & 2.7793 & 2.0747 & 0.7465 & 0.444 $\pm$ 0.004 & 0.5645 \\
    \bottomrule
  \end{tabular}
  \label{tab:drop}
\end{table}

\begin{table}[H]
  \centering
  \caption{Drop impact force and dissipation metrics.}
  \begin{tabular}{@{}lll@{}}
    \toprule
    Specimen & First impact force (N) & $E_{\mathrm{acc}}$ (mJ) \\
    \midrule
    LHR & 44.4 & 10.36 $\pm$ 0.70 \\
    MHR & 48.7 & 2.21 $\pm$ 0.45 \\
    HHR & 59.1 & 1.67 $\pm$ 0.16 \\
    \bottomrule
  \end{tabular}
  \label{tab:dynamic}
\end{table}

% Please provide Biographies and photos for Essays, Feature Articles, Progress Reports, Reviews, and Perspectives for those authors who should be highlighted.
% These should be at most 100 words long.
% Photographs should be 40 mm broad and 50 mm high.

%\begin{figure}[H]
%  \includegraphics{bio-placeholder.jpg}
%  \caption*{Biography}
%\end{figure}

% Table of contents entry should be 50 - 60 words long.
% Image should be 55 mm broad and 50 mm high or 110 mm broad and 20 mm high.
\begin{figure}[H]
\centering
\textbf{Table of Contents}

\medskip

\includegraphics[width=0.55\linewidth, trim={0.0cm 11.0cm 0.0cm 0.0cm}]{fig1.pdf}

\captionsetup{width=0.55\linewidth}
\caption*{\footnotesize
A kirigami meta-sheet is proposed as a planar impact absorber that uses transitions between positive and negative stiffness regimes instead of sacrificial crushing. Programming the hinge ratio connecting the unit cells allows the low hinge ratio meta-sheet to exhibit the clearest negative stiffness transition, reducing rebound, lowering first impact force, and increasing dissipation. The planar sheet also protects a falling egg, demonstrating practical impact absorption in thin structures.}
\end{figure}

\clearpage


\section*{Supplementary material}

Supporting Information

# Kirigami Meta-Sheet for Enhanced Impact Absorption

*Dahyun Joo, Yasuhiro Miyazawa, Seokhyeon Hong, Dongha Lee, Jinkyu Yang and Do-Nyun Kim**

This PDF file includes:

Figure S1 to S4

Table S1 to S4

Movie S1 to S3

References

**Supplementary Note 1. Fabrication of kirigami meta-sheet specimens**

Kirigami meta-sheet specimens were fabricated using an Ultimaker S5 fused filament fabrication 3D printer. Arnitel ID2045, also referred to as Addigy F2045, was selected for the compliant kirigami sheet because its elastomeric response allows large deformation of the hinge regions without brittle fracture. The softness of the filament, however, made the printing process sensitive to the Bowden-type feeding system of the Ultimaker S5. The filament behaved as a compliant spring inside the Bowden tube rather than as a rigid rod. Consequently, excessive feeder tension flattened the filament, increased tube friction, and promoted grinding. Short start-stop motions in the infill region further produced delayed pressure transfer and extrusion instability.

The final print protocol was therefore designed to maintain continuous nozzle pressure rather than to compensate for under-extrusion by globally increasing temperature or flow. The validated baseline used the recovered Arnitel ID2045 material profile, the weakest feeder grip that still provided stable feeding, a clean and straight Bowden path, disabled retraction, and slow continuous infill motion. If material feeding failed, the nozzle was recovered by selecting the machine's filament-change workflow and performing a hot-pull/cold-pull cleaning step using polylactic acid (PLA) before reloading Arnitel ID2045.

**Table S1:** Validated baseline parameters for printing Arnitel ID2045 kirigami meta-sheets with an Ultimaker S5.

| Category | Parameter | Baseline value |
|---|---|---|
| Material | Filament | Arnitel ID2045 |
| Hardware | Printer | Ultimaker S5, Bowden system |
| Hardware | Feeder tension | Weakest grip that still feeds without sliding |
| Hardware | Bowden path | Clean, straight, and re-seated or replaced when worn |
| Slicer | Nozzle temperature | 250 °C |
| Slicer | Overall flow / infill flow / top-bottom flow | 100% / 100% / 100% |
| Slicer | Retraction | Off |
| Slicer | Wall speed / infill speed / travel speed | $4\,\mathrm{mm.s^{-1}}$ / $5\,\mathrm{mm.s^{-1}}$ / $150\,\mathrm{mm.s^{-1}}$ |
| Slicer | Acceleration | $3800\,\mathrm{mm.s^{-2}}$ |
| Slicer | Horizontal expansion | 0 |
| Slicer | Infill path | Lines, continuous paths |
| Geometry | STL mesh | Densified when hinge quality was poor |

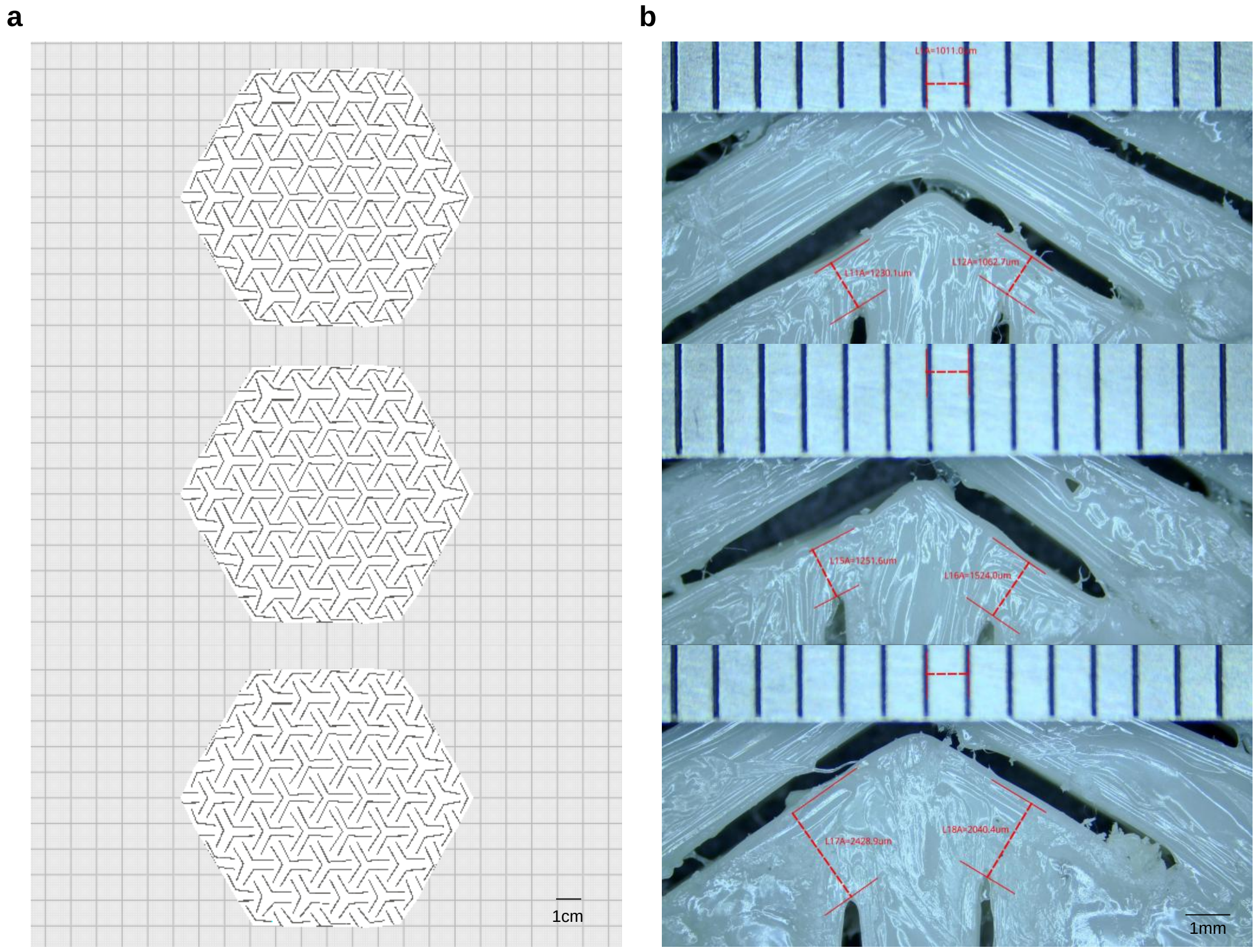


**Figure S1.** Fabrication layout and hinge ratio calibration of kirigami meta-sheet specimens. (a) Ultimaker print-screen view of the kirigami meta-sheet geometry. One grid cell corresponds to 1 cm. (b) LUSIS AUTO FOCUS CAMERA microscope images used to measure the as fabricated hinge thickness used for hinge ratio calibration.

## Supplementary Note 2. Hinge ratio measurement and ABAQUS indentation simulation setup

The planar kirigami pattern was generated using the previously reported computational design procedure for deployable bistable kirigami surfaces with finite hinge widths [1]. In the present study, the designed pattern was printed as a flat meta-sheet and subsequently fixed to a reinforced frame to reproduce the boundary condition used in finite element simulations.

The meta-sheet design was first verified on the Ultimaker print screen, where one grid cell corresponds to 1 cm (Figure S1a). The nominal hinge ratio (HR) values in the design file were 0.06250, 0.09000, and 0.15000 for the low HR (LHR), medium HR (MHR), and high HR (HHR) configurations, respectively. Manufacturing deviations can arise during printing, especially at the narrow flexible hinges. Therefore, the as fabricated hinge dimensions were measured using a LUSIS AUTO FOCUS CAMERA, and the HR used in the ABAQUS model was back-calculated from the actual specimen geometry rather than from the nominal computer

aided design (CAD) value (Figure S1b). The characteristic normalization length for this design is the effective unit cell length, $L_{\mathrm{uc}} = 14.25\,\mathrm{mm}$. This value corresponds to $L/8$ because eight effective unit cells span the longest dimension $L$ of the present kirigami pattern; therefore, the measured hinge thickness and the dimensionless HR are reported separately.

For each specimen, representative hinge thicknesses near the center of the meta-sheet were measured from microscope images. The HR was calculated by normalizing the measured hinge thickness, $t_h$, by the effective unit cell length, $L_{\mathrm{uc}}$:

$$HR = \frac{t_h}{L_{\mathrm{uc}}} \tag{S1}$$

consistent with the HR definition used in the hinge-controlled kirigami framework [1].

The mean measured hinge thicknesses and calibrated hinge ratios are summarized in Table S2. These measured values were used directly in the ABAQUS simulations so that the numerical models reflected the as fabricated specimens.

**Table S2:** Mean measured hinge thicknesses and calibrated HR values used in the finite element models.

| Specimen | Hinge thickness $t_h$ (mm) | HR |
|---|---|---|
| LHR | 1.1464 | 0.08045 |
| MHR | 1.3878 | 0.09739 |
| HHR | 2.2347 | 0.15682 |

The ABAQUS models used in the main text were adapted from the previously reported computational framework for hinge-controlled kirigami surfaces [1]. To improve numerical efficiency, the dome-forming meta-sheet was modeled using one sixth rotational symmetry. Symmetry boundary conditions were applied on the radial symmetry edges using local cylindrical coordinate systems, and the centerline was fully constrained.

The compliant sheet was modeled as a nearly incompressible Mooney–Rivlin hyperelastic material with strain-energy density

$$W = C_{10}(\bar{I}_1 - 3) + C_{01}(\bar{I}_2 - 3) + \frac{1}{D_1}(J-1)^2, \tag{S2}$$

where $C_{10} = 1.85472$ MPa, $C_{01} = 5.15168$ MPa, and $D_1 = 5.7088 \times 10^{-5}$ MPa$^{-1}$. The density was set to $730\,\mathrm{kg.m^{-3}}$ ($7.3 \times 10^{-10}$ tonne mm$^{-3}$). Stiffness-proportional Rayleigh damping with $\beta = 0.008\,\mathrm{s}$ and mass-proportional damping $\alpha = 0$ s$^{-1}$ was used to suppress high-frequency numerical noise. Self-contact and chuck–sheet contact were modeled using hard

normal contact with a penalty formulation and Coulomb friction with coefficient $\mu = 0.3$.
The mesh mainly used hybrid eight-node brick elements (C3D8H) to reduce volumetric locking in the nearly incompressible hyperelastic material. Six-node wedge elements (C3D6) and four-node tetrahedral elements (C3D4) were used where required by geometry. Five elements were assigned through the thickness direction. The hinge region was seeded with eight elements, and the long and short edges of the rigid-unit region were seeded with 12 and 4 elements, respectively, with a base element size of approximately 0.15 mm. This mesh density was selected to resolve large deformation bending and hinge localized stress concentration.
For the quasistatic indentation simulations, Dynamic Implicit displacement control was used with the same moderate-dissipation Hilber–Hughes–Taylor (HHT) setting ($\alpha = -0.05$) and geometrical nonlinearity enabled. The indentation speed was fixed at 80 mm.min$^{-1}$ ($\approx$ 1.333 mm.s$^{-1}$). To compare stress fields under the same deformation level, all three HR models were indented to 25 mm from the contact position. The von Mises stress field at 25 mm indentation was used for the stress comparison in the main text.

**Table S3:** Key ABAQUS model and analysis settings used for quasistatic indentation simulations.

| Category | Setting |
|---|---|
| Symmetry and boundary conditions | One-sixth rotational model; radial symmetry boundary conditions and fully constrained centerline |
| Material model | Mooney–Rivlin hyperelasticity with $C_{10} = 1.85472$ MPa, $C_{01} = 5.15168$ MPa, $D_1 = 5.7088 \times 10^{-5}$ MPa$^{-1}$; density 730 kg.m$^{-3}$ |
| Contact and damping | Hard normal contact, Coulomb friction $\mu = 0.3$, Rayleigh damping $\alpha = 0$ s$^{-1}$ and $\beta = 0.008$ s |
| Mesh | C3D8H-dominant mesh with C3D6/C3D4 where required; five through-thickness layers; hinge seed 8; base element size $\approx$ 0.15 mm |
| Indentation | Dynamic Implicit displacement control at 80 mm.min$^{-1}$; common target depth 25 mm for LHR, MHR, and HHR stress comparison |

### Supplementary Note 3. Reinforced PLA frame and specimen assembly

A PLA Orange frame was fabricated using the same Ultimaker S5 printer to hold the kirigami meta-sheet during indentation and drop impact experiments. PLA was selected for the frame because, among the readily accessible printable materials in the laboratory, it provided a practical balance of printability, stiffness, and resistance to parasitic bending. The frame was designed to constrain the perimeter of the meta-sheet while leaving the deformable central kirigami region

exposed to the indenter or impact chuck. A frame following only the outer boundary was not sufficiently rigid: during indentation or impact, the PLA frame itself bent noticeably, which made it difficult to isolate the mechanical response of the meta-sheet. Reinforcing ribs were therefore added to the side regions of the frame to reduce parasitic frame bending while leaving accessible tying points for the fishing lines.

The reinforced frame design was prepared on the Ultimaker screen, where one grid cell corresponds to 1 cm (Figure S2a). After printing, the meta-sheets were tightly fixed to the PLA frames using fishing lines, and adhesive was applied to lock the tied boundary during indentation and drop impact experiments (Figure S2b). The assembled specimens correspond to the LHR, MHR, and HHR meta-sheets from left to right (Figure S2b).

a

b

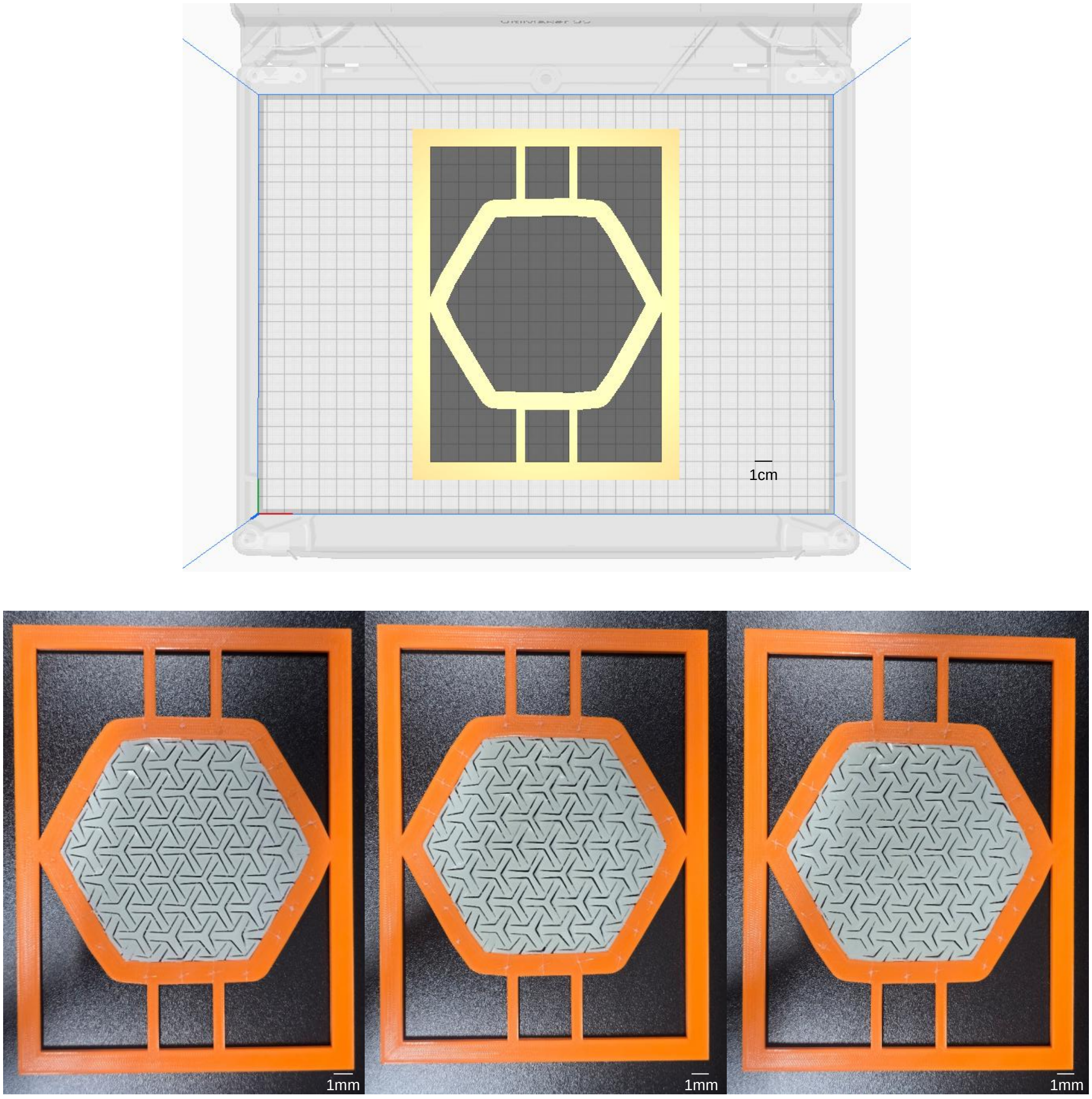


**Figure S2.** Reinforced PLA frame and assembled meta-sheet specimens. (a) Ultimaker printscreen view of the reinforced PLA Orange frame. One grid cell corresponds to 1 cm. (b) Fabricated specimens tightly fixed by fishing lines and adhesive for indentation and drop impact experiments. Left, LHR; middle, MHR; right, HHR.

## Supplementary Note 4. Indenter and drop-tower chuck fabrication

PLA chucks were fabricated for both quasistatic indentation and dynamic drop impact experiments. The contact diameter was 30 mm in both cases to maintain a consistent contact geometry between the two experimental protocols. The indentation chuck was printed as a single cylindrical part for mounting to the universal tensile machine. The drop-test chuck was separated into female and male components so that it could be assembled into the drop-tower system and mechanically fixed using screws.

The indentation chuck was printed as a single 30 mm-diameter PLA part for direct mounting to the universal tensile machine (Figure S3a,c). The drop-test chuck was printed as male and female 30 mm-diameter parts and assembled with screws for attachment to the drop-tower carriage (Figure S3b,d).

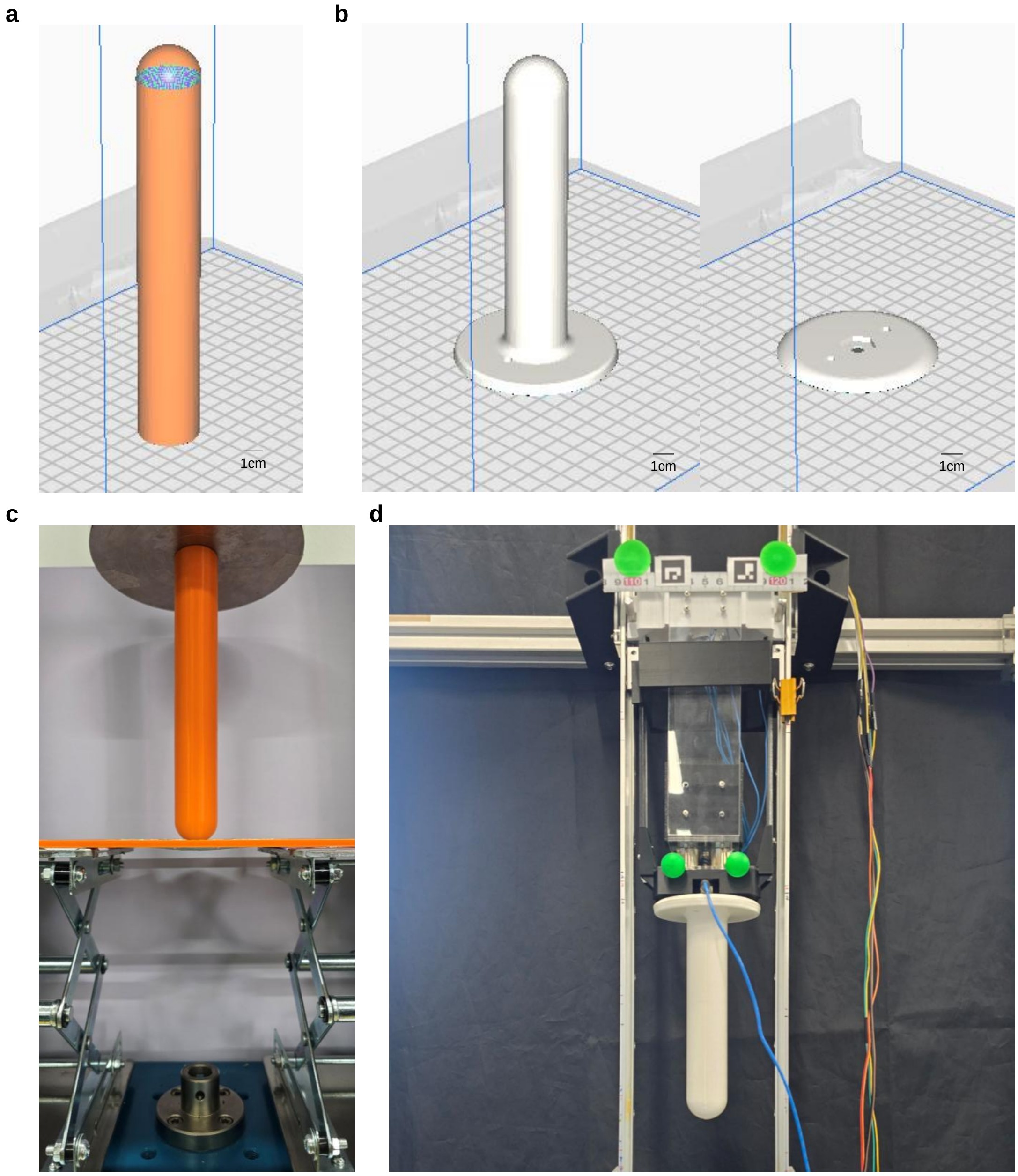

**Figure S3.** PLA chucks used for indentation and drop impact experiments. (a) Printing layout of the 30 mm-diameter indentation chuck. (b) Printing layout of the male and female parts of the 30 mm-diameter drop-test chuck. (c) Indentation chuck mounted to the universal tensile machine. (d) Drop-test chuck mounted to the drop-tower system.

**Supplementary Note 5. Experimental setup and instrumentation**

In the quasistatic indentation setup, the specimen was placed on two laboratory lifts, and the heights of the lifts were adjusted using a spirit level before each test so that the frame remained horizontal (Figure S4a). The 30 mm-diameter PLA indentation chuck was attached to the universal tensile machine and brought into vertical contact with the central region of the meta-sheet.

In the drop impact setup, the carriage holding the PLA chuck was fixed by an electromagnet while current was supplied (Figure S4b). When the switch was turned off, the current was interrupted, the electromagnet released the carriage, and the chuck fell onto the specimen. High speed video recording and acceleration measurement were triggered together with the impact event. A CHRONOS 1.4 high speed camera was used to record the drop-tower impact, and a SONY RX10 III camera was additionally used to record the kirigami deformation during both indentation and impact tests. The local surface acceleration of the kirigami meta-sheet was measured using a Bruel & Kjaer Charge Accelerometer Type 4393. The accelerometer was bonded to the middle of a boomerang-shaped connection between neighboring triangular units, at a position one unit cell away from the impact point, so that the sensor captured the vibration transmitted from the impact region without interfering with the direct contact area. The accelerometer signal was amplified and recorded using an oscilloscope. According to the calibration chart supplied with the sensor, the reference charge sensitivity was $0.3206\,\mathrm{pC/(m.s^{-2})}$, equivalent to $3.144\,\mathrm{pC/g}$, at 159.2 Hz. The amplifier setting used for post-processing converted voltage to acceleration using $10\,\mathrm{mV/(m.s^{-2})}$.

a b

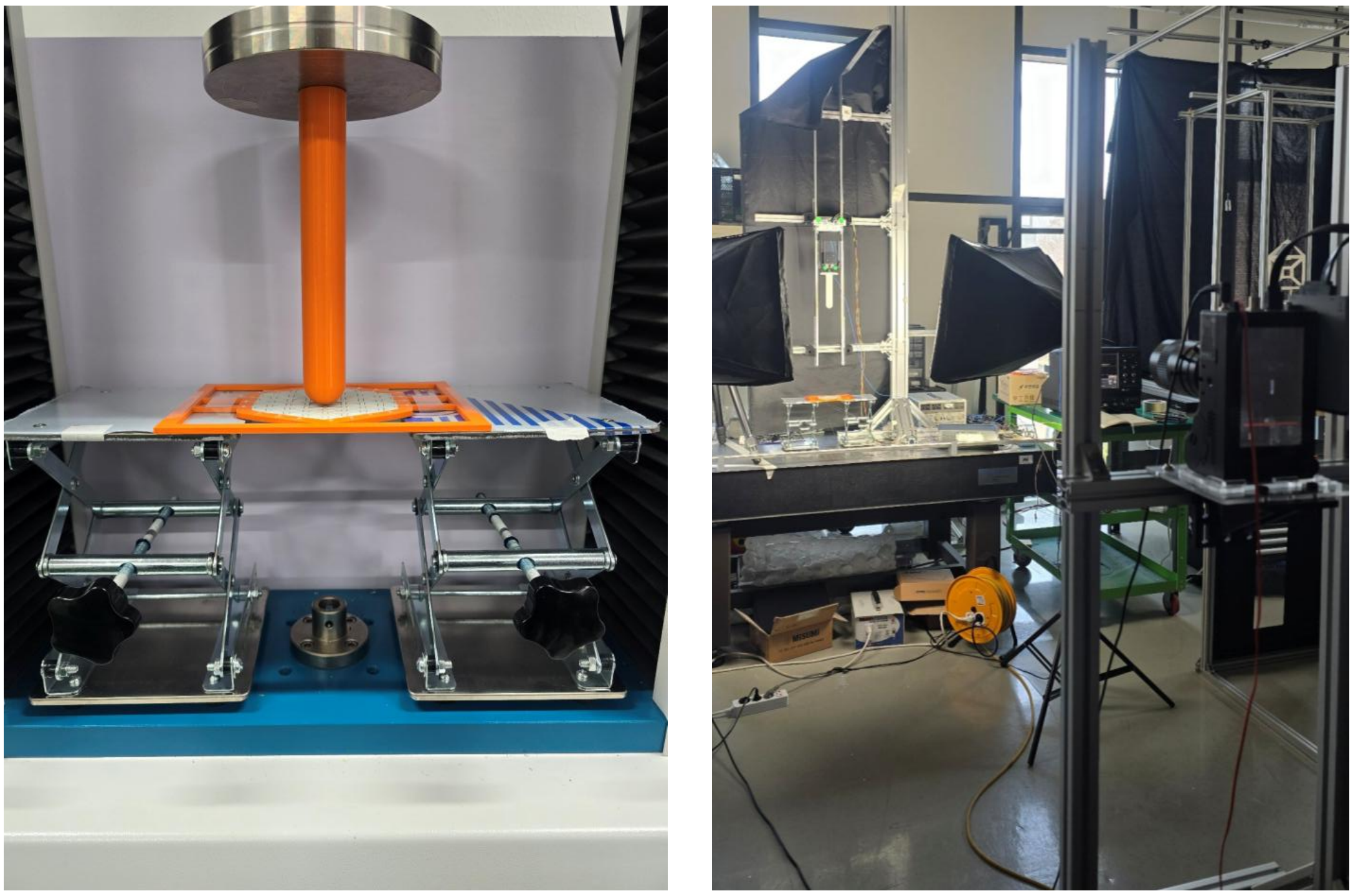

**Figure S4.** Experimental setups. (a) Quasistatic indentation setup with the specimen supported by two leveled laboratory lifts. (b) Drop-tower setup for synchronized impact, high speed video recording, and accelerometer measurement.

**Supplementary Note 6. High speed video time calibration**

The raw displacement trajectories extracted from the high speed videos contained a time-axis error caused by an incorrect frame-rate mapping in the tracking software. The nominal tracking time step was 0.00033 s, corresponding to 3000 frames per second (fps). However, the free fall segment of the drop motion produced an apparent acceleration substantially larger than gravitational acceleration, indicating that the video time axis had been compressed.

The time-axis correction was obtained from a first-principles free fall constraint. For a freely falling object over a distance $h = 0.428\,\mathrm{m}$, the physical fall time should satisfy

$$t_{\mathrm{true}} = \sqrt{\frac{2h}{g}} \tag{S3}$$

where $g = 9.80665\,\mathrm{m.s^{-2}}$. This gives $t_{\mathrm{true}} \approx 0.295\,\mathrm{s}$, whereas the raw tracking data gave $t_{\mathrm{meas}} \approx 0.179\,\mathrm{s}$. To calibrate the full time axis, the pre-impact free fall region was fitted using a quadratic form,

$$z(t_{\mathrm{meas}}) = p_0 t_{\mathrm{meas}}^2 + p_1 t_{\mathrm{meas}} + p_2 \tag{S4}$$

and the apparent acceleration was calculated as

$$a_{\text{meas}} = 2|p_0| \tag{S5}$$

If the corrected time is $t_{\text{real}} = Kt_{\text{meas}}$, the acceleration scales as $a_{\text{true}} = a_{\text{meas}}/K^2$. Therefore, the time-calibration factor is

$$K = \sqrt{\frac{a_{\text{meas}}}{g}} \tag{S6}$$

The calibrated value was $K = 1.4359$. Thus, the video time stamps were corrected as

$$t_{\text{real}} = 1.4359\, t_{\text{meas}} \tag{S7}$$

and the effective physical frame rate was

$$FPS_{\text{actual}} = \frac{FPS_{\text{measured}}}{K} = \frac{3000}{1.4359} = 2089.3 \text{ fps} \tag{S8}$$

This correction was applied only to video-derived tracking data. The accelerometer data were acquired using the hardware clock of the data-acquisition system and were therefore retained at their original sampling interval, $\Delta t = 0.001$ s.

**Supplementary Note 7. Video-derived velocity and impact velocity**

The corrected video displacement data were used to estimate the impact velocity immediately before contact. Direct finite-difference differentiation amplifies pixel-level tracking noise. To reduce this numerical noise without suppressing the physically sharp velocity peak before impact, the corrected displacement trajectory $z(t)$ was smoothed using a Savitzky-Golay filter with a 51-frame window. The velocity was then computed from the smoothed trajectory:

$$v(t_i) = \left. \frac{dz_{\text{SG}}(t)}{dt} \right|_{t=t_i} \tag{S9}$$

The impact velocity $v_0$ for each trial was defined as the maximum downward velocity immediately before contact:

$$v_0 = \max_{t<t_{\text{impact}}} |v(t)| \tag{S10}$$

The average calibrated impact velocities for the three specimen groups are listed in Table S4. These values include the effects of rail friction and small aerodynamic losses and were used for

rebound and damping calibration.

**Table S4:** Calibrated dynamic impact velocities obtained from corrected high speed video tracking. Impact velocities were extracted from the $K$-calibrated height trajectory using a Savitzky–Golay filter with a 51-frame window.

| Specimen group | Mean impact velocity $v_0$ (m s$^{-1}$) |
|---|---|
| LHR | 2.77 |
| MHR | 2.77 |
| HHR | 2.78 |

**Supplementary Note 8. Accelerometer calibration and voltage to acceleration conversion**

The same accelerometer measurement protocol was used consistently for the LHR, MHR, and HHR groups to measure the local vertical acceleration of the kirigami surface. The raw signal was recorded as an amplified voltage, $V(t)$. The amplifier output sensitivity was set to

$$S_{\text{acc}} = 10\,\text{mV}/(\text{m.s}^{-2}) = 0.01\,\text{V}/(\text{m.s}^{-2}) \tag{S11}$$

The raw accelerometer voltage contained a small hardware baseline offset of approximately $-50\,\text{mV}$, which caused the uncorrected acceleration trace to begin away from zero. To remove this bias, the pre-impact baseline was computed independently for each trial using the first 100 data points:

$$V_{\text{bias}} = \frac{1}{N_0}\sum_{i=1}^{N_0} V(t_i), \qquad N_0 = 100 \tag{S12}$$

The physical acceleration was then reconstructed as

$$a(t) = \frac{V(t) - V_{\text{bias}}}{S_{\text{acc}}} \tag{S13}$$

When the fixed nominal offset $V_{\text{bias}} \approx -0.05\,\text{V}$ is used, the acceleration-conversion expression becomes

$$a(t) = 100\,[V(t) + 0.05] \quad [\text{m}\,\text{s}^{-2}] \tag{S14}$$

This baseline correction enforces a common zero-acceleration reference before impact and prevents small amplifier offsets from being interpreted as physical acceleration.

**Supplementary Note 9. Filtering and time alignment of acceleration data**

The corrected acceleration traces were processed to suppress integration drift and to allow direct comparison across repeated trials. Small direct-current (DC) offsets and very low-frequency drift components can accumulate during integration and generate non-physical velocity drift. A first-order Butterworth high pass filter with cutoff frequency $f_c = 2.0\,\text{Hz}$ was therefore applied to the acceleration data:

$$a_{\text{HP}}(t) = \mathscr{H}_{\text{BW}}\left[a(t); f_c = 2.0\,\text{Hz}, n = 1\right] \tag{S15}$$

where $\mathscr{H}_{\text{BW}}$ denotes the Butterworth high pass filtering operator and $n = 1$ is the filter order. The resulting acceleration can be integrated to obtain velocity and displacement changes over the impact interval:

$$v(t) = v(t_0) + \int_{t_0}^{t} a_{\text{HP}}(\tau)\,d\tau \tag{S16}$$

$$u(t) = u(t_0) + \int_{t_0}^{t} v(\tau)\,d\tau \tag{S17}$$

For repeated trials, the impact time $t_{\text{impact}}$ was defined as the instant at which the acceleration began to increase sharply from the pre-impact baseline. To compare five repeated trials on the same time axis, all traces were shifted so that the plotting window began 0.24 s before $t_{\text{impact}}$:

$$t_{\text{aligned}} = t - \left(t_{\text{impact}} - 0.24\,\text{s}\right) \tag{S18}$$

A total time window of 3.0 s was used for plotting so that the post-impact vibration and subsequent settling process were both included.

## Supplementary Movies

Movie S1: Indentation experiments of LHR, MHR, and HHR kirigami meta-sheets. For each specimen group, the front-view indentation recording is followed by the corresponding bottom-view recording.

Movie S2: Drop-tower impact experiments of LHR, MHR, and HHR kirigami meta-sheets. For each specimen group, the drop impact recording is followed by the high speed surface-deformation recording.

Movie S3: Egg drop experiments comparing falling-egg protection by the larger kirigami meta-sheet, polyethylene (PE) mesh, and styrofoam control absorbers.



\end{document}